\documentclass[letterpaper,english,american,amsmath,amssymb,showpacs,superscriptaddress]{revtex4}
\usepackage[T1]{fontenc}
\usepackage{fancyhdr}
\usepackage{babel}
\usepackage{prettyref}
\usepackage{amstext}
\usepackage{graphicx}
\usepackage[unicode=true,
 bookmarks=true,bookmarksnumbered=false,bookmarksopen=false,
 breaklinks=false,pdfborder={0 0 1},backref=false,colorlinks=false]
 {hyperref}
\hypersetup{pdftitle={Reduction of colored noise in excitable systems to white noise and dynamic boundary conditions},
 pdfauthor={Schuecker, Diesmann, Helias},
 pdfsubject={Manuscript for arxiv},
 pdfkeywords={Transfer function, LIF neuron},
 colorlinks=true,linkcolor=black,citecolor=black,urlcolor=black,filecolor=black}
\usepackage{breakurl}

\makeatletter

\@ifundefined{textcolor}{}
{%
 \definecolor{BLACK}{gray}{0}
 \definecolor{WHITE}{gray}{1}
 \definecolor{RED}{rgb}{1,0,0}
 \definecolor{GREEN}{rgb}{0,1,0}
 \definecolor{BLUE}{rgb}{0,0,1}
 \definecolor{CYAN}{cmyk}{1,0,0,0}
 \definecolor{MAGENTA}{cmyk}{0,1,0,0}
 \definecolor{YELLOW}{cmyk}{0,0,1,0}
}

\newrefformat{cap}{\hyperref[#1]{Figure~\ref{#1}}}
\newrefformat{fig}{\hyperref[#1]{Figure~\ref{#1}}}
\newrefformat{tab}{\hyperref[#1]{Table ~\ref{#1}}}
\newrefformat{sec}{\hyperref[#1]{Section~\ref{#1}}}
\newrefformat{sub}{\hyperref[#1]{Section~\ref{#1}}}
\newrefformat{cha}{\hyperref[#1]{Chapter~\ref{#1}}}
\newrefformat{app}{\hyperref[#1]{Appendix~\ref{#1}}}

\makeatletter
\renewcommand{\p@subsection}{}
\renewcommand{\p@subsubsection}{}
\makeatother

\newcommand{\citebrunel}{\text{\citet{Brunel01_2186}}}
\newcommand{\citeklosek}{\text{\citet{Klosek98}}}

\makeatother

\begin{document}
\global\long\def\erf{\mathrm{erf}}
\global\long\def\LN{\mathcal{L}_{0}}
\global\long\def\LO{\mathcal{L}_{1}}
\global\long\def\mV{\:\mathrm{mV}}
\global\long\def\ms{\:\mathrm{ms}}
\global\long\def\Hz{\:\mathrm{Hz}}

\title{Reduction of colored noise in excitable systems to white noise and
dynamic boundary conditions}

\author{Jannis Schuecker }

\affiliation{Institute of Neuroscience and Medicine (INM-6) and Institute for
Advanced Simulation (IAS-6) and JARA BRAIN Institute I, Jülich Research
Centre, Jülich, Germany}

\author{Markus Diesmann }

\affiliation{Institute of Neuroscience and Medicine (INM-6) and Institute for
Advanced Simulation (IAS-6) and JARA BRAIN Institute I, Jülich Research
Centre, Jülich, Germany}

\affiliation{Department of Psychiatry, Psychotherapy and Psychosomatics, Medical
Faculty, RWTH Aachen University, Aachen, Germany}

\affiliation{Department of Physics, Faculty 1, RWTH Aachen University, Aachen,
Germany}

\author{Moritz Helias}

\affiliation{Institute of Neuroscience and Medicine (INM-6) and Institute for
Advanced Simulation (IAS-6) and JARA BRAIN Institute I, Jülich Research
Centre, Jülich, Germany}

\affiliation{Department of Physics, Faculty 1, RWTH Aachen University, Aachen,
Germany}

\date{\today}

\begin{abstract}
A recent study on the effect of colored driving noise on the escape
from a metastable state derives an analytic expression of the transfer
function of the leaky integrate-and-fire neuron model subject to colored
noise. Here we present an alternative derivation of the results, taking
into account time-dependent boundary conditions explicitly. This systematic
approach may facilitate future extensions beyond first order perturbation
theory. The analogy of the quantum harmonic oscillator to the LIF
neuron model subject to white noise enables a derivation of the well
known transfer function simpler than the original approach. We offer
a pedagogical presentation including all intermediate steps of the
calculations.
\end{abstract}
\maketitle

\section{Introduction}

In a recent study \citep{Schuecker14_arxiv_1411} we show that the
effect of colored noise on the escape from a metastable state can
be captured by a parametric shift of the location of the boundary
conditions for the probability density in the corresponding effective
system driven by white noise. We offer an alternative view of the
effective white-noise system, explicitly using a time-dependent boundary
condition at the original location. To linear order in the perturbation
parameter $k$ (cf. \prettyref{eq:diffeq_general}) these two approaches
are identical. While the shift of the boundary location is generic
and applicable to any arbitrary system at hand for which the white
noise solution is available, the approach given here comes along with
additional calculations. In the application to the leaky integrate-and-fire
(LIF) model neuron we first include a complete and simplified derivation
of the white-noise transfer function \citep{Brunel99,Lindner01_2934}
and subsequently present the colored noise calculations in the alternative
view.

\section{Effective diffusion \label{sec:Effective-diffusion}}

\subsection{Effective diffusion: a heuristic argument}

Consider the pair of coupled stochastic differential equations (SDE)
with a slow component $y$ with time scale $\tau$ driven by a fast
Ornstein-Uhlenbeck process $z$ with time scale $\tau_{s}$. In dimensionless
time $s=t/\tau$ and with $k=\sqrt{\tau_{s}/\tau}$ relating the two
scales we have

\begin{eqnarray}
\partial_{s}y & = & f(y,s)+\frac{z}{k}\label{eq:diffeq_general}\\
k\partial_{s}z & = & -\frac{z}{k}+\xi,\nonumber 
\end{eqnarray}
with a unit variance white noise $\langle\xi(s+u)\,\xi(s)\rangle=\delta(u)$.
We are interested in the case $\tau_{s}\ll\tau$ and start with a
heuristic argument on how to map the system of coupled SDEs to a single
diffusion equation. The subsequent sections will detail this mapping.
The autocorrelation function of $z$ is (see \prettyref{app:auto1d})

\begin{eqnarray*}
\langle z(s)z(s+s^{\prime})\rangle & = & \frac{1}{2}e^{-|s^{\prime}|/k^{2}}
\end{eqnarray*}
with time scale $k^{2}\propto\tau_{s}$. Since $y$ integrates $z$
on a time scale $\tau\gg\tau_{s}$, the effective quantity determining
the variance of $y$ is the integral of the autocorrelation function
of $z$ 

\begin{eqnarray}
\int\frac{1}{2}e^{-|s|/k^{2}}ds & = & k^{2}.\label{eq:integral_auto_2d}
\end{eqnarray}
We can compare this result to the limit $k\rightarrow0$, i.e. the
adiabatic approximation of \prettyref{eq:diffeq_general}, where
$z(s)$ follows $\xi(s)$ instantaneously. Thus $z(s)=k\xi(s)$ becomes
a white noise with autocorrelation $k^{2}\delta(s)$, yielding the
same integral of the autocorrelation as for finite $\tau_{s}$ \prettyref{eq:integral_auto_2d}.
Therefore the slow component $y$ effectively obeys the one-dimensional
SDE $\partial_{s}y=f(y,s)+\xi(s)$.

\subsection{Effective diffusion: a formal derivation\label{sub:formal derivation}}

We will now formalize the preceding heuristic argument. In order
to derive an effective one-dimensional diffusion equation for the
component $y$ and to obtain a formulation in which we can include
the treatment of absorbing boundary conditions, we consider the Fokker-Planck
equation \citep{Risken96} corresponding to \prettyref{eq:diffeq_general}
\begin{eqnarray}
k^{2}\partial_{s}P & = & \partial_{z}\left(\frac{1}{2}\partial_{z}+z\right)\,P-k^{2}\partial_{y}S_{y}\,P,\label{eq:FP_2D}
\end{eqnarray}
where $P(y,z,s)$ denotes the probability density and 
\begin{eqnarray}
S_{y} & = & f(y,s)+\frac{z}{k}\label{eq:flux_y_direction}
\end{eqnarray}
is the probability flux in $y$-direction. In order to obtain a perturbation
expansion in terms of simple eigenfunctions of the $z$-dependent
fast part of the Fokker-Planck operator, we factor-off its stationary
solution so that $P(y,z,s)=Q(y,z,s)\,\frac{e^{-z^{2}}}{\sqrt{\pi}}$.
Inserting the product into \prettyref{eq:FP_2D}, the chain rule suggests
the definition of a new differential operator $L$ acting on $Q$
by observing 
\begin{eqnarray}
\partial_{z}e^{-z^{2}}\circ & = & e^{-z^{2}}(\partial_{z}-2z)\circ\nonumber \\
\text{and}\nonumber \\
\left(\frac{1}{2}\partial_{z}+z\right)e^{-z^{2}}\circ & = & e^{-z^{2}}\left(\frac{1}{2}\partial_{z}-z+z\right)\circ=e^{-z^{2}}\frac{1}{2}\partial_{z}\circ\label{eq:transform_L}\\
\text{so compactly}\nonumber \\
\partial_{z}\left(\frac{1}{2}\partial_{z}+z\right)e^{-z^{2}}\circ & = & e^{-z^{2}}\,\left(\frac{1}{2}\partial_{z}-z\right)\partial_{z}\circ\equiv e^{-z^{2}}L\circ.\nonumber 
\end{eqnarray}
Expressed in $L$ the Fokker-Planck equation \prettyref{eq:FP_2D}
transforms to

\begin{eqnarray}
k^{2}\partial_{s}Q & = & LQ-kz\partial_{y}Q-k^{2}\partial_{y}\,f(y,s)\,Q.\label{eq:FP_Q_general}
\end{eqnarray}
In the following we refer to $Q$ as the outer solution, since initially
we do not consider the boundary conditions. We aim at an effective
Fokker-Planck equation for the $z$-marginalized solution $\tilde{P}(y,s)=\int dz\,\frac{e^{-z^{2}}}{\sqrt{\pi}}Q(y,z,s)$
that is correct up to linear order in $k$. This is equivalent to
knowing the first order correction to the marginalized probability
flux $\nu_{y}(y,s)\equiv\int dz\,\frac{e^{-z^{2}}}{\sqrt{\pi}}S_{y}Q(y,z,s)$.
Due to the form of \prettyref{eq:flux_y_direction} this requires
calculation of $Q$ up to second order in $k$. In addition we keep
only those terms that contribute to the zeroth and first order of
\foreignlanguage{english}{$\nu_{y}(y,s)$}. Inserting the perturbation
ansatz 
\begin{eqnarray}
Q(y,z,s) & = & \sum_{n=0}^{2}k^{n}\,Q^{(n)}(y,z,s)+O(k^{3})\label{eq:perturbation_expansion_k}
\end{eqnarray}
 into \prettyref{eq:FP_Q_general} we obtain 
\begin{eqnarray}
LQ^{(0)} & = & 0\label{eq:perturb_expansion_general}\\
LQ^{(1)} & = & z\partial_{y}Q^{(0)}\nonumber \\
LQ^{(2)} & = & \partial_{s}Q^{(0)}+z\partial_{y}Q^{(1)}+\partial_{y}fQ^{(0)}.\nonumber 
\end{eqnarray}
Noting the property $Lz^{n}=\frac{1}{2}n(n-1)z^{n-2}-nz^{n}$, we
see that the lowest order does not imply any further constraints on
the $z$-independent solution $Q^{(0)}(y,s)$, which must be consistent
with the solution to the one-dimensional Fokker-Planck equation corresponding
to the limit $k\to0$ of \prettyref{eq:diffeq_general}. With $Lz=-z$
the particular solution for the first order is 
\begin{eqnarray}
Q^{(1)}(y,z,s) & = & Q_{0}^{(1)}(y,s)-z\partial_{y}Q^{(0)}(y,s),\label{eq:Q1_general}
\end{eqnarray}
where we have the freedom to choose a function $Q_{0}^{(1)}(y,s)$
so far not constrained further except being independent of $z$, due
to $L\,1=0$. To generate the term linear in $z$ on the right hand
side of the second order in \prettyref{eq:perturb_expansion_general},
we need a term $-z\partial_{y}Q^{(1)}$. The terms constant in $z$
require contributions proportional to $z^{2}$, because $Lz^{2}=-2z^{2}+1$.
However, they can be dropped right away, because terms $\propto k^{2}z^{2}$
only contribute to the correction of the flux in order of $k^{2}$,
while their contribution to the first order resulting from the application
of the $\propto k^{-1}$ term in \prettyref{eq:flux_y_direction}
vanishes after marginalization. For the same reason the homogeneous
solution $Q_{0}^{(2)}(y,s)$ can be dropped. Hence the relevant part
of the second order solution is

\begin{eqnarray*}
Q^{(2)}(y,z,s) & = & -z\partial_{y}Q^{(1)}(y,s)+\text{terms causing \ensuremath{\nu_{y}\propto O(k^{2})}}.
\end{eqnarray*}
Inserting \prettyref{eq:Q1_general}, we also omit the term $z^{2}\partial_{y}^{2}Q^{(0)}$
as it is again $\propto k^{2}z^{2}$ and are left with
\begin{eqnarray}
Q(y,z,s) & = & Q^{(0)}(y,s)+kQ_{0}^{(1)}(y,s)-kz\partial_{y}Q^{(0)}(y,s)-k^{2}z\partial_{y}Q_{0}^{(1)}(y,s)+\text{terms causing \ensuremath{\nu_{y}\propto O(k^{2})}.}\label{eq:Q_eff}
\end{eqnarray}
Calculating the resulting flux marginalized over the fast variable
$z$ amounts to deriving the form of the effective flux operator acting
on the slow, $y$-dependent component. With \prettyref{eq:Q_eff}
this results in
\begin{eqnarray}
\nu_{y}(y,s) & = & \int dz\,\frac{e^{-z^{2}}}{\sqrt{\pi}}S_{y}Q\nonumber \\
 & = & \left(f(y,s)-\frac{1}{2}\partial_{y}\right)\left(Q^{(0)}(y,s)+kQ_{0}^{(1)}(y,s)\right)+O(k^{2}),\label{eq:effective_flux}
\end{eqnarray}
where we used $\int dz\,\frac{e^{-z^{2}}}{\sqrt{\pi}}=1$ and $\int dz\,\frac{z^{2}e^{-z^{2}}}{\sqrt{\pi}}=\frac{1}{2}$.
We recognize that $f(y,s)-\frac{1}{2}\partial_{y}$ is the flux operator
of a one-dimensional system driven by unit variance white noise and 

\begin{eqnarray}
\tilde{P}(y,s) & \equiv & Q^{(0)}(y,s)+kQ_{0}^{(1)}(y,s)\label{eq:marginalized_density}
\end{eqnarray}
corresponds to the marginalization of the relevant terms in \prettyref{eq:Q_eff}
over $z$, whereby the terms linear in $z$ vanish. Note that in \prettyref{eq:Q_eff}
the higher order terms in $k$ appear due to the operator $kz\partial_{y}$
in \prettyref{eq:FP_Q_general} that couples the $z$ and $y$ coordinate.
Eq. \prettyref{eq:effective_flux} shows that these terms cause an
effective flux that only depends on the $z$-marginalized solution
$\tilde{P}(y,s)$. This allows us to obtain the time evolution by
applying the continuity equation to the effective flux \prettyref{eq:effective_flux}
yielding the effective Fokker-Planck equation 
\begin{eqnarray}
\partial_{s}\tilde{P} & = & -\partial_{y}\nu_{y}(y,s)\nonumber \\
 & = & \partial_{y}\left(-f(y,s)+\frac{1}{2}\partial_{y}\right)\,\tilde{P}.\label{eq:FP_tilde}
\end{eqnarray}
For specific boundary conditions given by the physics of the system,
we need to determine the corresponding boundary condition for the
marginalized density. This amounts to determining the boundary condition
for $Q_{0}^{(1)}$, because we assume the one-dimensional white noise
problem to be exactly solvable and hence the boundary value of $Q^{(0)}$
to be known. Effective Fokker-Planck equations have been derived earlier
\citep{Sancho82,Lindenberg83,Haenggi85,Fox86,Grigolini86}, but these
approaches have been criticized for lacking a proper treatment of
the boundary conditions \citep{Klosek98}. In the framework introduced
in \citep{Doering87,Klosek98,Fourcaud02}, boundary conditions are
deduced using boundary layer theory for the two-dimensional Fokker-Planck
equation. In the next section we extend this framework to the transient
case.

\subsection{Effective boundary conditions\label{sub:Effective-boundary}}

For the dynamics \prettyref{eq:diffeq_general} with an absorbing
boundary at $y=\theta$, the flux vanishes for all points $\left(\theta,z\right)$
along the border with negative velocity in $y$-direction; these are
given by $f(\theta,s)+\frac{z}{k}<0$. Thus, the boundary condition
lives on a half line in $y,z$-space. This suggests, a translation
of the $z$ coordinate to
\begin{eqnarray}
z+kf(\theta,s) & \to & z,\label{eq:shifted_z}
\end{eqnarray}
so that $S_{y}=f(y,s)-f(\theta,s)+\frac{z}{k}$ is the flux operator
in $y$-direction \prettyref{eq:flux_y_direction} in the new coordinate
$z$. The boundary condition at threshold then takes the form
\begin{eqnarray}
0 & = & \frac{z}{k}\,Q(\theta,z,s)\quad\forall z<0\label{eq:flux_threshold}
\end{eqnarray}
and it follows that 
\begin{eqnarray}
Q(\theta,z,s) & = & 0\quad\forall z<0.\label{eq:Q_threshold_boundary_layer}
\end{eqnarray}
If, after absorption by the boundary, the system is reset to a smaller
value by assigning $y\leftarrow R$, this corresponds to the flux
escaping at threshold being re-inserted at reset. The corresponding
boundary condition is
\begin{eqnarray*}
\nu_{y}(\theta,z,s) & = & S_{y}Q(\theta,z,s)=S_{y}\left(Q(R+,z,s)-Q(R-,z,s)\right).
\end{eqnarray*}
With \prettyref{eq:flux_threshold} it follows that
\begin{eqnarray*}
\left(f(R,z,s)-f(\theta,z,s)+\frac{z}{k}\right)\left(Q(R+,z,s)-Q(R-,z,s)\right) & = & 0\quad\forall z<0,
\end{eqnarray*}
from which we conclude
\begin{eqnarray}
0= & Q(R+,z,s)-Q(R-,z,s) & \quad\forall z<0.\label{eq:Q_reset_boundary_layer}
\end{eqnarray}

This boundary conditions allows a non-continuous marginalized solution
at reset and enables us to deduce the value for the jump of the marginalized
density \prettyref{eq:marginalized_density}. Due to the time dependence
of the coordinate $z$ \prettyref{eq:shifted_z} the Fokker-Planck
equation \prettyref{eq:FP_2D} with the time derivative of the density
$\partial_{s}P(y,z(s),s)=\partial_{s}P-k(\partial_{s}f(\theta,s))\,\partial_{z}P$
transforms to

\begin{eqnarray}
k^{2}\partial_{s}P & = & \partial_{z}(\frac{1}{2}\partial_{z}+z-kf(\theta,s))\,P-k^{2}\partial_{y}(f(y,s)-f(\theta,s)+\frac{z}{k})P-k^{3}\,\partial_{s}f(\theta,s)\partial_{z}P.\label{eq:FP_time_depending_P}
\end{eqnarray}
With $P=Q\,\frac{e^{-z^{2}}}{\sqrt{\pi}}$ we obtain\foreignlanguage{english}{
}
\begin{eqnarray}
k^{2}\partial_{s}Q & = & LQ-kz\partial_{y}Q\label{eq:FP_outer_Q}\\
 &  & +k\left[f(\theta,s)(2z-\partial_{z})-k\partial_{y}\,(f(y,s)-f(\theta,s))\right]Q\nonumber \\
 &  & +k^{3}\,(\partial_{s}f(\theta,s))(2z-\partial_{z})Q.\nonumber 
\end{eqnarray}
The last term originating from the time dependence of $f$ is of third
order in $k$ and will therefore be neglected in the following. To
derive the boundary condition for the effective diffusion, we need
to describe the behavior of the original system near these boundaries
by transforming to either of the two shifted and scaled coordinates
$r=\frac{y-\{\theta,R\}}{k}$. To treat the reset condition analogously
to the condition at threshold, we introduce two auxiliary functions
$Q^{+}$ and $Q^{-}$: here $Q^{+}$ is a continuous solution of \prettyref{eq:FP_outer_Q}
on the whole domain and, above reset, agrees to the solution that
obeys the boundary condition at reset. Correspondingly, the continuous
solution $Q^{-}$ agrees to the searched-for solution below reset.
Due to linearity of \prettyref{eq:FP_outer_Q} also 
\begin{eqnarray*}
Q^{B}(r,z,s) & \equiv & \begin{cases}
Q(y(r),z,s) & \quad\text{at threshold}\\
Q^{+}(y(r),z,s)-Q^{-}(y(r),z,s)\equiv\Delta Q(y(r),z,s) & \quad\text{at reset}
\end{cases}
\end{eqnarray*}
is a solution. With this definition, the two boundary conditions \prettyref{eq:Q_threshold_boundary_layer}
and \prettyref{eq:Q_reset_boundary_layer} take the same form $Q^{B}(0,z,s)=0\quad\forall z<0.$
The coordinate $r$ zooms into the region near the boundary and changes
the order in $k$ of the interaction term $-kz\partial_{y}Q$ between
the $y$ and the $z$ component from first to zeroth order, namely
\begin{eqnarray}
k^{2}\partial_{s}Q^{B} & = & LQ^{B}-z\partial_{r}Q^{B}\nonumber \\
 &  & +k\left[f(\theta,s)(2z-\partial_{z})-\partial_{r}\,(f(kr+\{\theta,R\},s)-f(\theta,s))\right]\,Q^{B}+O(k^{3}).\label{eq:FP_boundary_variable}
\end{eqnarray}
With a perturbation ansatz in $k$, i.e. $Q^{B}=\sum_{n=0}^{1}k^{n}Q^{B(n)}+O(k^{2})$,
we obtain
\begin{eqnarray}
LQ^{B(0)}-z\partial_{r}Q^{B(0)} & = & 0\label{eq:order0_boundary}\\
LQ^{B(1)}-z\partial_{r}Q^{B(1)} & = & \left[f(\theta,s)(2z-\partial_{z})-\partial_{r}\,(f(kr+\{\theta,R\},s)-f(\theta,s))\right]\,Q^{B(0)}.\label{eq:order1_boundary}
\end{eqnarray}
The boundary layer solution must match the outer solution. Since the
outer solution varies only weakly on the length scale of $r$, a first
order Taylor expansion of the outer solution at the boundary yields
the matching condition

\begin{eqnarray}
Q^{B}(r,z,s) & = & \begin{cases}
Q(\theta,z,s)+kr\,\partial_{y}Q(\theta,z,s) & \text{at threshold}\\
\Delta Q(R,z,s)+kr\,\partial_{y}\Delta Q(R,z,s) & \text{at reset}.
\end{cases}\label{eq:QT_boundary}
\end{eqnarray}
To zeroth order in $k$ we hence have 

\begin{eqnarray}
Q^{B(0)}(0,z,s) & = & 0,\label{eq:QT0_boundary}
\end{eqnarray}
because the white noise system with $k=0$ has a vanishing density
at threshold and is continuous at reset. Together with the homogeneous
partial differential equation \prettyref{eq:order0_boundary} this
implies $Q^{B(0)}=0$ everywhere. To perform the matching of the first
order of \prettyref{eq:QT_boundary} we need to express the outer
solution in the shifted coordinate $z$ \prettyref{eq:shifted_z}.
The first order of the perturbation expansion \prettyref{eq:perturbation_expansion_k}
of the outer solution expressed in the new coordinate \prettyref{eq:shifted_z}
has a vanishing correction term $f(\theta,s)\frac{\partial Q^{(0)}(y,s)}{\partial z}=0$.
We can therefore insert \prettyref{eq:Q1_general} into \prettyref{eq:QT_boundary}
to obtain

\begin{eqnarray}
Q^{B(1)}(r,z,s) & = & \begin{cases}
Q_{0}^{(1)}(\theta,s)-z\partial_{y}Q^{(0)}(\theta,s)+r\partial_{y}Q^{(0)}(\theta,s) & \quad\text{at threshold}\\
\Delta Q_{0}^{(1)}(R,s)-z\partial_{y}\Delta Q^{(0)}(R,s)+r\partial_{y}\Delta Q^{(0)}(R,s) & \quad\text{at reset.}
\end{cases}\label{eq:QT1_boundary_general}
\end{eqnarray}
At threshold \prettyref{eq:QT1_boundary_general} can be simplified
to

\begin{eqnarray}
Q^{B(1)}(r,z,s) & = & Q_{0}^{(1)}(\theta,s)+2\nu_{y}^{(0)}(s)\,(z-r),\label{eq:QT1_boundary}
\end{eqnarray}
where we again exploit that $Q^{(0)}(\theta,s)=0$ in the white noise
system and therefore 
\begin{eqnarray}
\partial_{y}Q^{(0)}(\theta,s) & = & -2(-\frac{1}{2}\partial_{y}+f(\theta,s))\,Q^{(0)}(\theta,s)=-2\nu_{y}^{(0)}(\theta,s).\label{eq:flux_white_noise}
\end{eqnarray}
. Here $\nu_{y}^{(0)}(\theta,s)$ is the instantaneous flux at the
boundary of the white noise system. At reset \prettyref{eq:QT1_boundary_general}
takes the form

\begin{eqnarray}
Q^{B(1)}(r,z,s) & = & \Delta Q_{0}^{(1)}(R,s)+2\nu_{y}^{(0)}(\theta,s)\,(z-r),\label{eq:QT1_reset}
\end{eqnarray}
where we again use the continuity $\Delta Q^{(0)}(R,s)=\left.Q^{(0)}(y,s)\right|_{y=R-}^{y=R+}=0$
of the white noise system at reset and therefore \foreignlanguage{english}{$\partial_{y}\Delta Q^{(0)}(R,s)=\left.\partial_{y}Q^{(0)}(y,s)\right|_{y=R-}^{y=R+}=-2\nu_{y}^{(0)}(\theta,s)$}
.

\subsubsection{Half-range expansion}

Using \prettyref{eq:QT0_boundary} the first order solution \prettyref{eq:order1_boundary}
must satisfy 
\begin{eqnarray}
LQ^{B(1)}-z\partial_{r}Q^{B(1)} & = & 0.\label{eq:QT1}
\end{eqnarray}
With the definitions $v=\sqrt{2}r$, $w=\sqrt{2}z$, and $g(v,w,s)=Q^{B(1)}(r,z,s)$,
equation \prettyref{eq:QT1} takes the form
\begin{eqnarray*}
(\partial_{w}^{2}-w\partial_{w})\,g(v,w,s) & = & w\partial_{v}g(v,w,s).
\end{eqnarray*}
Note that the time argument plays the role of a parameter here, since
the time derivatives in \prettyref{eq:FP_outer_Q} and \eqref{eq:FP_boundary_variable}
are of higher order in $k$. With the absorbing boundary condition
$g(0,w,s)=0$ for $w<0$, following at threshold from \prettyref{eq:Q_threshold_boundary_layer}
and at reset from \prettyref{eq:Q_reset_boundary_layer}, the solution
not growing faster than linear in $v\to-\infty$ is given by \citet[B.11]{Klosek98}
\begin{eqnarray}
g(v,w,s) & = & \frac{C(s)}{\sqrt{2}}\,\left(\frac{\alpha}{\sqrt{2}}+w-v+\sqrt{2}\sum_{n=1}^{\infty}b_{n}(w/\sqrt{2})\,e^{\sqrt{n}v}\right)\label{eq:bl_solution}
\end{eqnarray}
with $\alpha=\sqrt{2}|\zeta(\frac{1}{2})|$ given by Riemann's $\zeta$-function
and $b_{n}$ proportional to the $n$-th Hermitian polynomial. The
constant $\alpha$ defined here follows the notation used in \citet{Fourcaud02}
and differs by a factor of $\sqrt{2}$ from the notation in \citet[B.11]{Klosek98}.
At threshold and in the original coordinates we equate \prettyref{eq:bl_solution}
to \prettyref{eq:QT1_boundary} which reads  
\begin{eqnarray*}
Q_{0}^{(1)}(y,s)+2\nu_{y}^{(0)}(\theta,s)\,(z-r) & = & C(s)\left(\frac{\alpha}{2}+z-r+\sum_{n=1}^{\infty}b_{n}(z)\,e^{\sqrt{2n}r}\right).
\end{eqnarray*}
The term proportional to $(z-r)$ fixes the time dependent function
$C(s)=2\nu_{y}^{(0)}(\theta,s)$. The exponential term on the right
hand side has no equivalent term on the left hand side. It varies
on a small length scale inside the boundary layer, while the terms
on the left hand side originate from the outer solution, varying on
a larger length scale. Therefore the exponential term can not be taken
into account and the term proportional to $\alpha$ fixes the boundary
value 
\begin{eqnarray}
Q_{0}^{(1)}(\theta,s) & = & \alpha\nu_{y}^{(0)}(\theta,s).\label{eq:Q_01_boundary_value}
\end{eqnarray}
At reset we equate \prettyref{eq:bl_solution} and \prettyref{eq:QT1_reset}
and consider the left-sided limit $y\uparrow R$ ensuring $r<0$,
which is sufficient to determine the value of $\Delta Q(R,s)$. We
obtain
\begin{eqnarray*}
\Delta Q_{0}^{(1)}(R,s)+2\nu_{y}^{(0)}(\theta,s)\,(z-r) & = & C(s)\left(\frac{\alpha}{2}+z-r+\sum_{n=1}^{\infty}b_{n}(z)\,e^{\sqrt{2n}r}\right),
\end{eqnarray*}
so that we find the jump of the outer solution at reset $\Delta Q_{0}^{(1)}(R,s)=\left.Q_{0}^{(1)}(y,s)\right|_{y=R-}^{y=R+}=\alpha\nu_{y}^{(0)}(\theta,s)$.
This concludes the central argument of the general theory: The effective
Fokker-Planck equation \prettyref{eq:FP_tilde} has the time-dependent
boundary conditions 
\begin{eqnarray}
\tilde{P}(\theta,s) & = & \underbrace{Q^{(0)}(\theta,s)}_{=0}+kQ_{0}^{(1)}(\theta,s)=k\alpha\nu_{y}^{(0)}(\theta,s)\label{eq:Boundary_conditions_static}\\
\left.\tilde{P}(y,s)\right|_{R-}^{R+} & = & \underbrace{\left.Q^{(0)}(y,s)\right|_{R-}^{R+}}_{=0}+k\left.Q_{0}^{(1)}(y,s)\right|_{R-}^{R+}=k\alpha\nu_{y}^{(0)}(\theta,s),\nonumber 
\end{eqnarray}
reducing the colored noise problem to the solution of a one-dimensional
Fokker-Planck equation, for which standard methods are available \citep{Risken96}.

\subsection{Shifted reset and threshold\label{sub:Shifted-reset-and}}

\citet[B.11]{Klosek98,Fourcaud02} stated that the steady state density
in the colored noise case can be interpreted as the solution to the
stationary white noise problem with shifted threshold $\tilde{\theta}=\theta+k\frac{\alpha}{2}$
and reset $\tilde{R}=R+k\frac{\alpha}{2}$. We formally show that
to first order in $k$ the dynamic boundary conditions \prettyref{eq:Boundary_conditions_static}
found for the time-dependent problem can as well be expressed as a
shift of the boundaries. We perform a Taylor expansion of the effective
density at $\tilde{\theta}$

\begin{eqnarray*}
\tilde{P}(\tilde{\theta},s) & = & \tilde{P}(\theta+k\frac{\alpha}{2},s)\\
 & = & \tilde{P}(\theta,s)+k\frac{\alpha}{2}\partial_{y}\tilde{P}(\theta,s)+O(k^{2})\\
 & \stackrel{\text{ \eqref{eq:Boundary_conditions_static} and \eqref{eq:flux_white_noise}}}{=} & k\alpha\nu_{y}^{(0)}(\theta,s)-k\frac{\alpha}{2}2\nu_{y}^{(0)}(\theta,s)+O(k^{2})\\
 & = & O(k^{2}),
\end{eqnarray*}
where in the second last step we used that the derivative $\partial_{y}\tilde{P}(\theta,s)$
with \prettyref{eq:flux_white_noise} can be written as $2\nu_{y}^{(0)}(\theta,s)-2f(\theta,s)\,\tilde{P}(\theta,s)=2\nu_{y}^{(0)}(\theta,s)+O(k)$.

For the reset we have 
\begin{eqnarray*}
\left.\tilde{P}(y,s)\right|_{\tilde{R}-}^{\tilde{R}+} & = & \left.\tilde{P}(y,s)\right|_{R-}^{R+}+k\frac{\alpha}{2}\partial_{y}\left.\tilde{P}(y,s)\right|_{R-}^{R+}+O(k^{2})\\
 & \stackrel{\eqref{eq:Boundary_conditions_static}}{=} & k\alpha\nu_{y}^{(0)}(\theta,s)+k\frac{\alpha}{2}\partial_{y}\left.Q^{(0)}(y,s)\right|_{R-}^{R+}+O(k^{2})\\
 & = & k\alpha\nu_{y}^{(0)}(\theta,s)+k\frac{\alpha}{2}\partial_{y}Q^{(0)}(\theta,s)+O(k^{2})\\
 & = & O(k^{2}),
\end{eqnarray*}
where we used that the white noise density is continuous at reset
and the flux at threshold $\nu_{y}^{(0)}(\theta,s)=-\frac{1}{2}\partial_{y}\,Q^{(0)}(\theta,s)$
is reinserted at reset, e.g $\partial_{y}\,Q^{(0)}(\theta,s)=\partial_{y}\left.Q^{(0)}(y,s)\right|_{R-}^{R+}$.
We conclude that the colored noise dynamic boundary conditions \prettyref{eq:Boundary_conditions_static}
are equivalent to white noise boundary conditions at shifted threshold
and reset. This view is pursued in \citep{Schuecker14_arxiv_1411},
while in the following sections we derive the colored-noise correction
explicitly using the time-dependent colored-noise boundary condition
\prettyref{eq:Boundary_conditions_static}.

\section{Example: LIF neuron model\label{sec:Example:-LIF-neuron}}

We now apply the general formalism to the leaky integrate-and-fire
(LIF) neuron model, revealing a novel analytical expression for the
transfer function for the case that the synaptic noise is filtered.
Previous work shows that this correction vanishes in the case of the
perfect integrate-and-fire model \citet[p. 2077]{Fourcaud02} in the
low frequency domain.

\subsection{The LIF model: the harmonic oscillator of neuroscience\label{sec:Leaky-integrate-and-fire-neuron-and-harmonic-oscillator}}

The membrane potential $V$ of the LIF neuron model without synaptic
filtering (white noise) evolves according to the differential equation

\begin{eqnarray}
\tau\dot{V} & = & -V+\mu+\sigma\sqrt{\tau}\xi(t),\label{eq:diff_LIF_wn}
\end{eqnarray}
where $\tau$ is the membrane time constant and the input is described
by mean $\mu$ and variance $\sigma^{2}$ in diffusion approximation.
If $V$ reaches the threshold $\theta$ the membrane potential is
reset to a smaller value $V\leftarrow R$. The corresponding Fokker-Planck
equation is
\begin{eqnarray}
\partial_{t}P(V,t) & = & -\partial_{V}\varphi(V,t)\nonumber \\
\varphi(V,t) & \equiv & \left(-\frac{1}{\tau}(V-\mu)-\frac{1}{2}\frac{\sigma^{2}}{\tau}\partial_{V}\right)P(V,t).\label{eq:FP_orig}
\end{eqnarray}
In dimensionless coordinates it takes the form
\begin{eqnarray}
\partial_{s}\rho(x,s) & = & -\partial_{x}\underbrace{(-x-\partial_{x})}_{\equiv S_{0}}\,\rho(x,s)\equiv\mathcal{L}_{0}\,\rho(x,s),\label{eq:FP_transformed}
\end{eqnarray}
where $x=\sqrt{2}\frac{V-\mu}{\sigma}$, $s=t/\tau$, \foreignlanguage{english}{$\rho(x,s)\equiv\frac{\sigma}{\sqrt{2}}P(V,t)$}
and $S_{0}$ is the probability flux operator. The Fokker-Planck operator
$\mathcal{L}_{0}$ is not Hermitian. However, we can transform the
operator to a Hermitian form, for which standard solutions are available.
We therefore follow \citet[p. 134, eq. 6.9]{Risken96} and apply a
transformation that is possible whenever the Fokker-Planck equation
possesses a stationary solution $\bar{\rho}_{0}$ (here $\bar{\rho}_{0}=e^{-\frac{x^{2}}{2}}$
is the stationary solution of \prettyref{eq:FP_transformed} if threshold
and reset are absent). We define a function $u(x)=e^{-\frac{1}{4}x^{2}}=\sqrt{\bar{\rho}_{0}(x)}$
and observe that it fulfills the following relations 
\begin{eqnarray}
\partial_{x}u(x)\circ & = & u(x)\underbrace{(-\frac{1}{2}x+\partial_{x})}_{\equiv-a^{\dagger}}\circ\label{eq:commutation_u}\\
(x+\partial_{x})u(x)\circ & = & u(x)\underbrace{(\frac{1}{2}x+\partial_{x})}_{\equiv a}\circ.\nonumber 
\end{eqnarray}
Here we defined the operators

\begin{eqnarray}
a & \equiv & \frac{1}{2}x+\partial_{x}\label{eq:operator_def}\\
a^{\dagger} & \equiv & \frac{1}{2}x-\partial_{x}=x-a\nonumber 
\end{eqnarray}
that fulfill the commutation relation
\begin{eqnarray}
[a,a^{\dagger}] & = & aa^{\dagger}-a^{\dagger}a=1.\label{eq:commutator}
\end{eqnarray}
Hence, writing $\rho(x,s)=u(x)\,q(x,s)$, the flux operator $S_{0}$
and the Fokker-Planck operator $\mathcal{L}_{0}$ transform to

\begin{eqnarray}
S_{0}\,u\circ=(-x-\partial_{x})\,u\circ & = & -ua\circ\label{eq:transformed_flux}\\
\mathcal{L}_{0}\,u\circ=\partial_{x}(x+\partial_{x})\,u\circ & = & -ua^{\dagger}a\circ.\nonumber 
\end{eqnarray}
The Fokker-Planck equation \prettyref{eq:FP_transformed} can then
be expressed in terms of $a^{\dagger}$ and $a$ as
\begin{eqnarray}
\partial_{s}q(x,s) & = & -a^{\dagger}a\,q(x,s)\label{eq:oscillator_basis_operator}\\
 & = & (\partial_{x}^{2}-\frac{1}{4}x^{2}+\frac{1}{2})\,q(x,s).\nonumber 
\end{eqnarray}
The right hand side is the Hamiltonian of the quantum harmonic oscillator.
Note, however, that the $i\hbar$ is missing on the left hand side.
The operator now is Hermitian and the eigenfunctions of $a^{\dagger}a$
form a complete orthogonal set. In the stationary case, the probability
flux between reset $x_{R}$ and threshold $x_{\theta}$ (with $x_{\{\theta,R\}}=\sqrt{2}\frac{\{\theta,R\}-\mu}{\sigma}$)
is constant ($\tau\nu_{0}$), whereas it vanishes below $x_{R}$ and
above $x_{\theta}$. With \prettyref{eq:transformed_flux} the flux
takes the form

\begin{eqnarray}
-ua\,q_{0} & = & \tau\nu_{0}\,H(x-x_{R})H(x_{\theta}-x).\label{eq:stationary_flux}
\end{eqnarray}
The homogeneous solution $a\,q_{h}=(\partial_{x}+\frac{1}{2}x)\,q_{h}(x)=0$
is $q_{h}(x)=u(x)$. Hence, the full solution satisfying the white
noise boundary condition $q_{0}(x_{\theta})=0$ is 
\begin{eqnarray}
q_{0}(x) & = & \tau\nu_{0}\,u(x)\,\int_{\max(x,x_{R})}^{x_{\theta}}u^{-1}(x^{\prime})u^{-1}(x^{\prime})\,dx^{\prime}\nonumber \\
 & = & \tau\nu_{0}\,e^{-\frac{x^{2}}{4}}\,\int_{\max(x,x_{R})}^{x_{\theta}}e^{\frac{x^{\prime2}}{2}}\,dx^{\prime}.\label{eq:density_unperturbed}
\end{eqnarray}
Consequently, the solution in terms of the density $\rho$ is $\rho_{0}(x)=u(x)\,q_{0}(x)=\tau\nu_{0}\,e^{-\frac{x^{2}}{2}}\,\int_{\max(x,x_{R})}^{x_{\theta}}e^{\frac{x^{\prime2}}{2}}\,dx^{\prime}$,
in agreement with \citep[eq. 19]{Brunel00_183}. We determine the
(as yet arbitrary) constant $\nu_{0}$ from the normalization condition
$1=\int\rho(x)\,dx$ as
\begin{align*}
(\tau\nu_{0})^{-1} & =\int_{-\infty}^{x_{\theta}}\rho(x)\,dx\\
 & =\int_{-\infty}^{x_{\theta}}e^{-\frac{x}{2}^{2}}\int_{\max(x,x_{R})}^{x_{\theta}}e^{\frac{x^{\prime2}}{2}}\,dx^{\prime}\,dx\\
 & =2\int_{-\infty}^{y_{\theta}}\underbrace{e^{-y^{2}}}_{f^{\prime}}\underbrace{\int_{\max(y,y_{R})}^{y_{\theta}}e^{u^{2}}\,du}_{g}\,dy,
\end{align*}
where in the last step we substituted $y=x/\sqrt{2}$ and $u=x^{\prime}/\sqrt{2}$.
Using integration by parts with $f(y)=\int_{-\infty}^{y}e^{-x^{2}}dx=\frac{\sqrt{\pi}}{2}(1+\erf(y))$
and $g^{\prime}(y)=-e^{y^{2}}H(y-y_{R})$ and noting that the boundary
term vanishes, because $g(y_{\theta})=0$ and $f(-\infty)=0$ we have
\begin{align}
(\tau\nu_{0})^{-1} & =\sqrt{\pi}\int_{y_{R}}^{y_{\theta}}(1+\erf(y))\,e^{y^{2}}\,dy.\label{eq:Siegert}
\end{align}
This is the formula originally found by \citeauthor{Siegert51} for
the mean-first-passage time determining the firing rate $\nu_{0}$
\citep{Siegert51,Brunel00_183}. The higher eigenfunctions of $\mathcal{L}_{0}u$
\prettyref{eq:transformed_flux} are obtained by repeated application
of $a^{\dagger}$, as the commutation relation $[a,a^{\dagger}]=1$
holds and hence 
\begin{eqnarray}
a^{\dagger}a(a^{\dagger})^{n}q_{0} & = & a^{\dagger}(a^{\dagger}a+[a,a^{\dagger}])(a^{\dagger})^{n-1}q_{0}\nonumber \\
 & = & a^{\dagger}(a^{\dagger}a+1)(a^{\dagger})^{n-1}q_{0}\nonumber \\
 & = & \ldots\nonumber \\
 & = & (a^{\dagger})^{n}(a^{\dagger}a+n)q_{0}=n(a^{\dagger})^{n}q_{0}.\label{eq:ladder_operators}
\end{eqnarray}
So the spectrum of the operator is discrete and specified by the set
of integer numbers $n\in\mathbb{N}_{0}.$

\subsection{Stationary firing rate for colored noise \label{sub:Stationary-firing-rate-cn}}

Let us now consider a leaky integrate-and-fire model neuron with synaptic
filtering, i.e. the system of coupled differential equations in diffusion
approximation \citep{Fourcaud02}

\begin{eqnarray}
\tau\dot{V} & = & -V+I+\mu\label{eq:diff_LIF}\\
\tau_{s}\dot{I} & = & -I+\sigma\sqrt{\tau}\xi(t).\nonumber 
\end{eqnarray}
The general system \prettyref{eq:diffeq_general} can be obtained
from \prettyref{eq:diff_LIF} by introducing the coordinates $z=\frac{k}{\sigma}I$,
setting $f(y,s)=-y$, and observing that the rescaling of the time
axis $s=t/\tau$ cancels a factor $\sqrt{\tau}$ in front of the noise,
because $\langle\sqrt{\tau}\xi(t+u)\sqrt{\tau}\xi(t)\rangle=\tau\delta(u)=\delta\left(\frac{u}{\tau}\right)=\langle\xi(s+\frac{u}{\tau})\xi(s)\rangle$.
The corresponding two-dimensional Fokker-Planck equation is \eqref{eq:FP_2D}
\begin{eqnarray}
k^{2}\partial_{s}P & = & \partial_{z}(\frac{1}{2}\partial_{z}+z)\,P-k^{2}\partial_{y}(-y+\frac{z}{k})\,P.\label{eq:time_dep_FP_LIF}
\end{eqnarray}
Using again $x=\sqrt{2}y$, and the marginalized density \foreignlanguage{english}{$\rho(x,s)=\frac{1}{\sqrt{2}}\tilde{P}(x/\sqrt{2},s),$}
the effective reduced system \prettyref{eq:FP_tilde} is the white
noise case \prettyref{eq:oscillator_basis_operator} with the boundary
conditions deduced from the half range expansion \prettyref{eq:Boundary_conditions_static}.
As in the white noise case \prettyref{eq:stationary_flux} we solve
\begin{eqnarray}
-ua\,q_{0}(x) & = & \tau\nu\,H(x-x_{R})H(x_{\theta}-x),\label{eq:stationary_flux-1}
\end{eqnarray}
where $\nu$ denotes the colored noise firing rate. With the homogeneous
solution $u$ the general solution is given by

\begin{eqnarray}
uq_{0} & = & \begin{cases}
uq_{+}\equiv D_{+}u^{2}+uq_{p} & \text{for\ }x_{R}<x<x_{\theta}\\
uq_{-}\equiv D_{-}u^{2} & \text{for}\ x<x_{R}
\end{cases}\label{eq:uq_CD}
\end{eqnarray}
with the particular solution of \prettyref{eq:stationary_flux-1}
for $x_{R}<x<x_{\theta}$ chosen to vanish at threshold 
\begin{eqnarray*}
uq_{p} & = & \tau\nu\,u^{2}(x)\int_{x}^{x_{\theta}}u{}^{-2}dx^{\prime}.
\end{eqnarray*}
The constants $D_{+}$ and $D_{-}$ are fixed by the boundary conditions
\prettyref{eq:Boundary_conditions_static}. With the stationary flux
$\nu_{y}^{(0)}=\tau\nu_{0}$ in the white noise system we get 

\begin{eqnarray}
u(x_{\theta})q_{+}(x_{\theta})=D_{+}u^{2}(x_{\theta}) & = & \frac{1}{\sqrt{2}}k\alpha\tau\nu_{0}\equiv A\label{eq:Boundary_values_C.1-1-1}\\
u(x_{R})q_{+}(x_{R})-u(x_{R})q_{-}(x_{R}) & = & A.\nonumber 
\end{eqnarray}
Thus we have $D_{+}=Au^{-2}(x_{\theta})$ and $D_{-}$ can be determined
as
\begin{eqnarray}
D_{+}u^{2}(x_{\theta})=A & = & u(x_{R})q_{+}(x_{R})-u(x_{R})q_{-}(x_{R})\nonumber \\
 & = & (D_{+}-D_{-})u^{2}(x_{R})+\tau\nu\,u^{2}(x_{R})\int_{x_{R}}^{x_{\theta}}u^{-2}\,dx\nonumber \\
\Leftrightarrow D_{-} & = & A\left(u^{-2}(x_{\theta})-u^{-2}(x_{R})\right)+\tau\nu\int_{x_{R}}^{x_{\theta}}u^{-2}\,dx.\label{eq:Constant D}
\end{eqnarray}
The firing rate $\nu$ is determined by the normalization condition
\begin{eqnarray}
1 & = & \int_{\infty}^{x_{\theta}}uq\,dx\label{eq:Normalization}\\
 & = & \int_{-\infty}^{x_{R}}uq_{-}\,dx+\int_{x_{R}}^{x_{\theta}}uq_{+}\,dx.\nonumber 
\end{eqnarray}
Inserting \prettyref{eq:uq_CD} with $D_{+}$ and $D_{-}$ suggests
the introduction of 
\begin{eqnarray*}
F(x) & = & \int_{-\infty}^{x}u^{2}\,dx^{\prime}\\
 & = & \int_{-\infty}^{x}e^{-\frac{1}{2}x^{\prime2}}\,dx^{\prime}\\
 & = & \sqrt{\frac{\pi}{2}}(1+\erf(\frac{x}{\sqrt{2}}))\\
\text{and}\\
I & = & \int_{x_{R}}^{x_{\theta}}u^{2}\int_{x}^{x_{\theta}}u^{-2}\,dx^{\prime}dx\\
 & \stackrel{\text{int. by parts}}{=} & -F(x_{R})\,\int_{x_{R}}^{x_{\theta}}u^{-2}\,dx^{\prime}+\int_{x_{R}}^{x_{\theta}}F(x)\,u^{-2}(x)\,dx.
\end{eqnarray*}
From \prettyref{eq:Normalization} we obtain

\begin{eqnarray*}
1 & = & A\left(u^{-2}(x_{\theta})-u^{-2}(x_{R})\right)\,F(x_{R})+Au^{-2}(x_{\theta})\,(F(x_{\theta})-F(x_{R}))+\tau\nu\,F(x_{R})\int_{x_{R}}^{x_{\theta}}u^{-2}\,dx+\tau\nu\,I\\
 & = & A\left.u^{-2}(x)F(x)\right|_{x_{R}}^{x_{\theta}}+\tau\nu\,\int_{x_{R}}^{x_{\theta}}u^{-2}(x)F(x)\,dx,
\end{eqnarray*}
so that

\begin{eqnarray*}
\tau\nu & = & \frac{1-A\left.u^{-2}F\right|_{x_{R}}^{x_{\theta}}}{\int_{x_{R}}^{x_{\theta}}u^{-2}F\,dx}.
\end{eqnarray*}
Furthermore the firing rate without synaptic filtering $\nu_{0}$
can be expressed as 

\begin{eqnarray}
\tau\nu_{0} & = & \frac{1}{\int_{x_{R}}^{x_{\theta}}u^{-2}F\,dx}.\label{eq:firing_rate_white_noise}
\end{eqnarray}
With \prettyref{eq:Boundary_values_C.1-1-1} we have 
\begin{eqnarray}
\tau\nu & = & \tau\nu_{0}-\tau\nu_{0}\,\frac{\alpha k}{\sqrt{2}}\,\frac{\left.u^{-2}F\right|_{x_{R}}^{x_{\theta}}}{\int_{x_{R}}^{x_{\theta}}u^{-2}F\,dx}\nonumber \\
 & = & \tau\nu_{0}-\frac{\alpha k}{\sqrt{2}}\frac{\left.u^{-2}F\right|_{x_{R}}^{x_{\theta}}}{\left(\int_{x_{R}}^{x_{\theta}}u^{-2}F\,dx\right)^{2}}\label{eq:firing_rate_final}
\end{eqnarray}
and finally determined the first order correction $\nu_{1}$ of the
perturbation ansatz $\nu=\nu_{0}+k\nu_{1}+O(k^{2})$ in agreement
with \citet{Fourcaud02}. Up to linear order in $k$ this is equivalent
to

\begin{eqnarray}
\tau\nu & = & \left(\int_{x_{R}+\frac{\alpha k}{\sqrt{2}}}^{x_{\theta}+\frac{\alpha k}{\sqrt{2}}}u^{-2}F\,dx\right)^{-1}\label{eq:shifted_bc}
\end{eqnarray}
as shown by Taylor expansion of the latter expression up to linear
order in $k$. Comparison of \prettyref{eq:shifted_bc} to the white
noise case \prettyref{eq:firing_rate_white_noise} shows that we can
reformulate \prettyref{eq:Boundary_conditions_static} as a shift
of the locations of the white noise boundaries by $\frac{\alpha k}{\sqrt{2}}$,
as found in \citet{Klosek98} for the static case.

\subsection{White noise transfer function\label{sub:White-noise-transfer-function}}

We now simplify the derivation of the transfer function of the LIF
neuron model for white noise \citep{Brunel99,Lindner01_2934} by exploiting
the analogy to the quantum harmonic oscillator introduced above. Consider
a periodic modulation of the mean input in \prettyref{eq:diff_LIF_wn}
\begin{eqnarray}
\mu(t) & = & \mu+\delta\mu(t)\label{eq:mean_modulation}\\
\delta\mu(t) & = & \epsilon\mu e^{i\omega t}\nonumber 
\end{eqnarray}
and the variance 

\begin{eqnarray*}
\sigma^{2}(t) & = & \sigma^{2}+\delta\sigma(t)^{2}\\
\delta\sigma(t)^{2} & = & H\sigma^{2}e^{i\omega t}.
\end{eqnarray*}
To linear order this will result in a modulation of the firing rate
$\nu_{0}(t)=\nu_{0}(1+n(\omega)e^{i\omega t})$, where $n(\omega)$
is the transfer function. Note that both modulations $\delta\mu$
and $\delta\sigma$ have their own contribution to $n(\omega)$ and
in principle could be treated separately since we only determine the
linear response here. For brevity we consider them simultaneously
here. The time dependent Fokker-Planck equation takes the form
\begin{eqnarray*}
\partial_{t}P(V,t) & = & -\partial_{V}(\varphi(V,t)+\delta\varphi(V,t))\\
\delta\varphi(V,t) & = & \left(\frac{\delta\mu(t)}{\tau}-\frac{\delta\sigma^{2}(t)}{2\tau}\partial_{V}\right)\,P(V,t)
\end{eqnarray*}
or, in the natural coordinates 

\begin{eqnarray}
\partial_{s}\rho(x,s) & = & \LN(x)\rho(x,s)\label{eq:perturbed_FP}\\
 & + & e^{i\omega\tau s}\underbrace{(-G\,\partial_{x}+H\partial_{x}^{2})}_{\equiv\LO(x)}\,\rho(x,s).\nonumber 
\end{eqnarray}
Here, $G=\sqrt{2}\epsilon\mu/\sigma$ and we defined the perturbation
operator $\LO(x)$.

\subsubsection{Perturbative treatment of the time-dependent Fokker-Planck equation\label{sub:Perturbative-treatment-of}}

For small amplitudes $n(\omega)\ll1$, so weak modulations of the
rate compared to the stationary baseline rate, we employ the ansatz
of a perturbation series, namely that the time-dependent solution
of \prettyref{eq:perturbed_FP} is in the vicinity of the stationary
solution, $\rho(x,s)=\rho_{0}(x)+\rho_{1}(x,s)$, with the correction
$\rho_{1}$ of linear order in the perturbing quantities $\delta\mu(t)$
and \foreignlanguage{english}{$\delta\sigma(t)$}. Inserting this
ansatz into \prettyref{eq:perturbed_FP} and using the property of
the stationary solution $\LN\rho_{0}=0$ we are left with an inhomogeneous
partial differential equation for the unknown function $\rho_{1}$
\begin{eqnarray}
\partial_{s}\rho_{1}(x,s) & = & \LN(x)\rho_{1}(x,s)+e^{i\omega\tau s}\,\LO(x)\rho_{0}(x)\label{eq:perturbation_FP}\\
 & + & e^{i\omega\tau s}\LO(x)\rho_{1}(x,s).\nonumber 
\end{eqnarray}
Neglecting the third term that is of second order in the perturbed
quantities, the separation ansatz $\rho_{1}(x,s)=\rho_{1}(x)\,e^{i\omega\tau s}$
(for brevity we drop the $\omega$-dependence of $\rho_{1}(x)$) then
leads to the linear ordinary inhomogeneous differential equation of
second order
\begin{eqnarray*}
i\omega\tau\,\rho_{1} & = & \LN\rho_{1}+\LO\rho_{0}.
\end{eqnarray*}
From here the operator representation introduced in \prettyref{sec:Leaky-integrate-and-fire-neuron-and-harmonic-oscillator}
guides us to the solution. Writing $\rho_{1}(x)=u(x)\,q_{1}(x)$ and
with the commutation relation \prettyref{eq:commutation_u} $\partial_{x}u(x)\circ=-u(x)\,a^{\dagger}\circ$
the transformed inhomogeneity takes the form `
\begin{eqnarray}
\LO u\,q_{0} & = & -\partial_{x}\underbrace{(G-H\,\partial_{x})}_{\equiv S_{1}}\,u\,q_{0}\nonumber \\
 & = & u\,a^{\dagger}(G+H\,a^{\dagger})\,q_{0},\label{eq:S1}
\end{eqnarray}
where we defined the contribution of the perturbation to the flux
operator as $S_{1}$. With $\LN uq_{1}=-ua^{\dagger}aq_{1}$ we need
to solve the equation
\begin{eqnarray}
(i\omega\tau+a^{\dagger}a)\,q_{1} & = & (G\,a^{\dagger}+H\,(a^{\dagger})^{2})\,q_{0}.\label{eq:inhomogene_fp_op}
\end{eqnarray}
Since the equation is linear in $q_{1}$, its solution is a superposition
of a particular solution and a homogeneous solution. The latter needs
to be chosen such that the full solution complies with the boundary
conditions but we first need to find the particular solution. To this
end we will use the property \prettyref{eq:ladder_operators}. For
$n=1$ and $n=2$ we have 
\begin{eqnarray*}
a^{\dagger}a\,(a^{\dagger}q_{0}) & = & a^{\dagger}q_{0}\\
\text{and }\\
a^{\dagger}a\,((a^{\dagger})^{2}q_{0}) & = & 2(a^{\dagger})^{2}q_{0}.
\end{eqnarray*}
Hence a term proportional to $a^{\dagger}q_{0}$ reproduces the first
part of the inhomogeneity in \prettyref{eq:inhomogene_fp_op} and
a term proportional to $(a^{\dagger})^{2}q_{0}$ generates the second
term. We therefore use $q_{p}=(\gamma a^{\dagger}+\beta(a^{\dagger})^{2})\,q_{0}$
as the ansatz for the particular solution and determine the coefficients
$\gamma$ and $\beta$ by inserting into \prettyref{eq:inhomogene_fp_op},
which yields
\begin{eqnarray*}
i\omega\tau\,(\gamma a^{\dagger}+\beta(a^{\dagger})^{2})\,q_{0}+(\gamma a^{\dagger}+2\beta(a^{\dagger})^{2})\,q_{0} & = & (G\,a^{\dagger}+H\,(a^{\dagger})^{2})\,q_{0}.
\end{eqnarray*}
Sorting by terms according to powers of $a^{\dagger}$, we obtain
two equations determining $\gamma,\beta$

\begin{eqnarray*}
(i\omega\tau\gamma+\gamma-G)\,a^{\dagger}q_{0} & = & 0\\
(i\omega\tau\beta+2\beta-H)\,(a^{\dagger})^{2}q_{0} & = & 0,
\end{eqnarray*}
where the factor in parenthesis must be nought, because neither $a^{\dagger}q_{0}(x)$
nor $(a^{\dagger})^{2}q_{0}(x)$ vanish for all $x$. This leaves
us with the particular solution
\begin{eqnarray}
q_{p} & = & \left(\frac{G}{1+i\omega\tau}\,a^{\dagger}q_{0}+\frac{H}{2+i\omega\tau}\,(a^{\dagger})^{2}q_{0}\right).\label{eq:particular_solution}
\end{eqnarray}
This equation together with \prettyref{eq:ladder_operators} shows
that the perturbed solution consists of the first and the second excited
state above the ground state, because the two terms are proportional
to $a^{\dagger}q_{0}$ and $\left(a^{\dagger}\right)^{2}q_{0}$. Thus
the modulation of the input to the neuron is equivalent to exciting
the harmonic oscillator to higher energy states. Since only the ground
state is a stationary solution, it is intuitively clear that the response
of the neuron relaxes back after some time, in analogy to the return
from the exited states.

\subsubsection{Homogeneous solution\label{sub:homogenous_solution}}

The homogeneous equation follows from \prettyref{eq:inhomogene_fp_op}

\begin{eqnarray}
(i\omega\tau+a^{\dagger}a)\,q_{h} & = & 0.\label{eq:parabolic_cylinder}
\end{eqnarray}
Evaluating $a^{\dagger}a$ yields 
\begin{eqnarray*}
(-\partial_{x}^{2}+\frac{1}{4}x^{2}+i\omega\tau-\frac{1}{2})\,q_{h} & = & 0,
\end{eqnarray*}
which can be rearranged to the form 

\begin{eqnarray}
\partial_{x}^{2}q_{h}-(\frac{1}{4}x^{2}+m)\,q_{h} & = & 0\label{eq:parabolic_cylinder_diffeq}\\
\text{with }\quad m & = & i\omega\tau-\frac{1}{2},\nonumber 
\end{eqnarray}
the solution of which can be written as a linear combination of two
parabolic cylinder functions \citep[12.2]{DLMF12}. The function
$U(m,x)=D_{-m-\frac{1}{2}}(x)$ of Whittaker \citep[19.3.1/2]{Abramowitz74}
has the asymptotic behavior $U(m,x)\propto e^{-\frac{1}{4}x^{2}}|x|^{-m-\frac{1}{2}}$
for $x\rightarrow-\infty$ \citep[19.8.1]{Abramowitz74}. The other
independent solution $V(m,x)\propto e^{\frac{1}{4}x^{2}}|x|^{m-\frac{1}{2}}$
is divergent for $|x|\to\infty$, so that $u(x)\,V(m,x)\propto|x|^{-1+i\omega\tau}$
is due to the logarithmic divergence not integrable on $(-\infty,0)$.
The contribution of $V(m,x)$ therefore needs to vanish in order
to arrive at a normalizable density. Due to the boundary conditions,
we distinguish two different domains 
\begin{eqnarray}
q_{1}(x) & = & q_{p}(x)+\begin{cases}
c_{1-}U(x) & \text{for }x<x_{R}\\
c_{1+}U(x)+c_{2+}V(x) & \text{for }x_{R}\le x<x_{\theta}.
\end{cases}\label{eq:general_solution_q1}
\end{eqnarray}
In the following we skip the dependence of the parabolic cylinder
function on $m$ for brevity of the notation. The homogeneous solution
(i.e. the coefficients $c_{1,2\pm}$) adjusts the complete solution
$q_{1}=q_{p}+q_{h}$ to the boundary conditions dictated by the physics
of the problem.

\subsubsection{Boundary condition for the modulated density}

The complete solution $q_{1}=q_{h}+q_{p}$ of \prettyref{eq:inhomogene_fp_op}
must fulfill the white noise boundary conditions

\begin{eqnarray}
q_{1}(x_{\theta}) & = & 0\quad\text{at threshold}\label{eq:finite_flux_threshold}\\
q_{1}(x_{R+})-q_{1}(x_{R-}) & = & 0\text{\quad at reset },\nonumber 
\end{eqnarray}
whereby it must vanish at threshold to ensure a finite probability
flux and be continuous at reset for the same reason. Introducing the
short hand 
\begin{eqnarray}
\left.f(x)\right|_{x^{\prime}} & = & \begin{cases}
f(x^{\prime}) & \text{for }x^{\prime}=x_{\theta}\\
f(x^{\prime}+)-f(x^{\prime}-) & \text{for }x^{\prime}=x_{R}
\end{cases}\label{eq:short_hand_x}
\end{eqnarray}
we state these two conditions compactly as
\begin{eqnarray*}
\left.q_{1}(x)\right|_{\{x_{R},x_{\theta}\}} & = & 0.
\end{eqnarray*}
To determine the boundary values of the homogeneous solution we need
the boundary values of the particular solution first. The latter follow
with \prettyref{eq:operator_def} and the stationary flux \prettyref{eq:stationary_flux}
\begin{eqnarray}
\left.a^{\dagger}q_{0}\right|_{\{x_{R},x_{\theta}\}} & \stackrel{\eqref{eq:operator_def}}{=} & \left.(x-a)\,q_{0}\right|_{\{x_{R},x_{\theta}\}}\stackrel{\eqref{eq:stationary_flux}}{=}u^{-1}(\{x_{R},x_{\theta}\})\,\tau\nu_{0},\label{eq:a_dagger_boundary}
\end{eqnarray}
where the term $xq_{0}$ vanishes because of the continuity of $q_{0}$.
Along the same lines follows the term proportional to $(a^{\dagger})^{2}q_{0}$
\begin{eqnarray*}
(a^{\dagger})^{2}q_{0} & = & (x-a)a^{\dagger}q_{0}\\
 & = & (xa^{\dagger}-[a,a^{\dagger}]-a^{\dagger}a)q_{0}\\
 & \stackrel{\eqref{eq:commutator}}{=} & (xa^{\dagger}-1)q_{0}.
\end{eqnarray*}
With \prettyref{eq:a_dagger_boundary} and the continuity of $q_{0}$
we therefore have

\begin{eqnarray}
\left.(a^{\dagger})^{2}q_{0}\right|_{\{x_{R},x_{\theta}\}} & = & \{x_{R},x_{\theta}\}u^{-1}(\{x_{R},x_{\theta}\})\,\tau\nu_{0}.\label{eq:a_ddagger_boundary}
\end{eqnarray}
From the explicit expression of the particular solution \prettyref{eq:particular_solution}
with the term proportional to $a^{\dagger}q_{0}$ specified by \prettyref{eq:a_dagger_boundary},
the term proportional to $(a^{\dagger})^{2}q_{0}$ by \prettyref{eq:a_ddagger_boundary}
and the continuity of the complete solution \prettyref{eq:finite_flux_threshold}
then follows the initial value for the homogeneous solution as

\begin{eqnarray}
-\left.q_{h}\right|_{\{x_{R},x_{\theta}\}} & = & \left.q_{p}\right|_{\{x_{R},x_{\theta}\}}\label{eq:threshold_cond}\\
 & = & \left(\frac{G}{1+i\omega\tau}+\frac{H}{2+i\omega\tau}\{x_{R},x_{\theta}\}\right)\,\tau\nu_{0}\,u^{-1}(\{x_{R},x_{\theta}\}).\nonumber 
\end{eqnarray}

\subsubsection{Boundary condition for the derivative of the density}

The boundary condition for the first derivative of $q_{1}$ follows
considering the probability flux: The flux at threshold must be equal
to the flux re-inserted at reset. Given the firing rate follows the
periodic modulation $\nu_{0}(t)=\nu_{0}(1+n(\omega)e^{i\omega t})$,
we can express the flux $\tau\nu_{0}\,n(\omega)$ due to the perturbation
(the stationary solution fulfills $\left.S_{0}uq_{0}\right|_{\{x_{R},x_{\theta}\}}=\left.-ua\,q_{0}\right|_{\{x_{R},x_{\theta}\}}=\tau\nu_{0}$)
as a sum of two contributions, corresponding to the first two terms
in \prettyref{eq:perturbation_FP}

\begin{eqnarray}
\tau\nu_{0}\,n(\omega) & = & \left.S_{0}uq_{1}+S_{1}uq_{0}\right|_{\{x_{R},x_{\theta}\}}\nonumber \\
 & = & \left.-ua\,q_{1}+u(G+Ha^{\dagger})q_{0}\right|_{\{x_{R},x_{\theta}\}}.\label{eq:flux_conditions}
\end{eqnarray}
Again we first evaluate the contribution of the particular solution
\prettyref{eq:particular_solution} considering

\begin{eqnarray*}
a\,a^{\dagger}q_{0} & \stackrel{\eqref{eq:commutator}}{=} & (1+a^{\dagger}a)q_{0}\\
 & = & q_{0}+\underbrace{a^{\dagger}aq_{0}}_{=0}.
\end{eqnarray*}
Analogously follows
\begin{eqnarray*}
a\,(a^{\dagger})^{2}q_{0} & = & 2a^{\dagger}q_{0},
\end{eqnarray*}
so that the flux due to the particular solution \prettyref{eq:particular_solution}
can be written as 
\begin{eqnarray}
-ua\,q_{p} & = & -u\left(\frac{G}{1+i\omega\tau}+\frac{2H}{2+i\omega\tau}\,a^{\dagger}\right)q_{0}.\label{eq:flux_contribution_particular}
\end{eqnarray}
As the stationary solution vanishes at threshold $q_{0}(x_{\theta})=0$
and is continuous at reset, the first term vanishes when inserted
into \prettyref{eq:flux_conditions}. Hence with \prettyref{eq:a_dagger_boundary}
the contribution to the flux \prettyref{eq:flux_conditions} yields
\begin{eqnarray*}
\left.-ua\,q_{p}\right|_{\{x_{R},x_{\theta}\}} & =- & \frac{2H\tau\nu_{0}}{2+i\omega\tau}.
\end{eqnarray*}
With \prettyref{eq:a_dagger_boundary} the term due to the perturbed
flux operator $S_{1}$ in \prettyref{eq:flux_conditions} is
\begin{eqnarray*}
\left.u\,(G+Ha^{\dagger})q_{0}\right|_{\{x_{R},x_{\theta}\}} & = & H\,\tau\nu_{0}.
\end{eqnarray*}
Inserting the previous two expressions into \prettyref{eq:flux_conditions}
we obtain
\begin{eqnarray}
\tau\nu_{0}\,n(\omega) & = & \tau\nu_{0}\frac{i\omega\tau H}{2+i\omega\tau}-\left.uaq_{h}\right|_{\{x_{R},x_{\theta}\}}\nonumber \\
\Leftrightarrow\left.u(\frac{1}{2}x+\partial_{x})q_{h}\right|_{\{x_{R},x_{\theta}\}} & = & \tau\nu_{0}\left(\frac{i\omega\tau H}{2+i\omega\tau}-n(\omega)\right),\label{eq:threshold_deriv_cond}
\end{eqnarray}
where we used the explicit form of $a=\frac{1}{2}x+\partial_{x}$.
The derivative then follows as
\begin{eqnarray}
\left.\partial_{x}q_{h}\right|_{\{x_{R},x_{\theta}\}} & = & \tau\nu_{0}\left(\frac{i\omega\tau H}{2+i\omega\tau}-n(\omega)\right)\,u^{-1}(\{x_{R},x_{\theta}\})-\left.\frac{1}{2}xq_{h}(x)\right|_{\{x_{R},x_{\theta}\}}\label{eq:deriv_qh}
\end{eqnarray}
and with \prettyref{eq:threshold_cond} we obtain
\begin{eqnarray}
\left.\partial_{x}q_{h}\right|_{\{x_{R},x_{\theta}\}} & = & \tau\nu_{0}\left(\frac{i\omega\tau H}{2+i\omega\tau}-n(\omega)+\frac{1}{2}\{x_{R},x_{\theta}\}\left(\frac{G}{1+i\omega\tau}+\frac{H}{2+i\omega\tau}\,\{x_{R},x_{\theta}\}\right)\right)\,u^{-1}(\{x_{R},x_{\theta}\})\label{eq:deriv_qh-1}\\
 & = & \tau\nu_{0}\left(-n(\omega)+\frac{1}{2}\{x_{R},x_{\theta}\}\frac{G}{1+i\omega\tau}+\left(\frac{1}{2}\{x_{R},x_{\theta}\}^{2}+i\omega\tau\right)\frac{H}{2+i\omega\tau}\right)\,u^{-1}(\{x_{R},x_{\theta}\}).\nonumber 
\end{eqnarray}

\subsubsection{Transfer function\label{sub:Solvability_delta}}

Having found the function value and the derivative at threshold, the
homogeneous solution (of the second order differential equation) is
uniquely determined on $x_{R}<x<x_{\theta}$. Writing the solution
on this interval as
\[
\begin{array}{ccccccc}
U(x_{\theta}) & c_{1+} & + & V(x_{\theta}) & c_{2+} & = & q_{1}^{h}(x_{\theta})\\
U^{\prime}(x_{\theta}) & c_{1+} & + & V^{\prime}(x_{\theta}) & c_{2+} & = & \partial_{x}q_{1}^{h}(x_{\theta}),
\end{array}
\]
the coefficients follow as the solution of this linear system of equations,
which is in matrix form 
\begin{eqnarray*}
\left(\begin{array}{cc}
U(x_{\theta}) & V(x_{\theta})\\
U^{\prime}(x_{\theta}) & V^{\prime}(x_{\theta})
\end{array}\right)\left(\begin{array}{c}
c_{1+}\\
c_{2+}
\end{array}\right) & = & \left(\begin{array}{c}
q_{h}(x_{\theta})\\
\partial_{x}q_{h}(x_{\theta})
\end{array}\right).
\end{eqnarray*}
The solution is

\begin{eqnarray}
\left(\begin{array}{c}
c_{1+}\\
c_{2+}
\end{array}\right) & = & \frac{1}{W(x_{\theta})}\left(\begin{array}{cc}
V^{\prime} & -V\\
-U^{\prime} & U
\end{array}\right)\left(\begin{array}{c}
q_{1}^{h}(x_{\theta})\\
\partial_{x}q_{1}^{h}(x_{\theta})
\end{array}\right)\label{eq:Solution_UVsystem}\\
\text{with}\nonumber \\
W & = & \det\left(\begin{array}{cc}
U & V\\
U^{\prime} & V^{\prime}
\end{array}\right),\nonumber 
\end{eqnarray}
where the function $W(x)$ is the Wronskian and for the given functions
$U,V$ is a constant $W=\sqrt{\frac{2}{\pi}}$ \citep[19.4.1]{Abramowitz74}.
The coefficients follow from the previous expression using \prettyref{eq:threshold_cond}
and \prettyref{eq:deriv_qh-1} and $c_{2+}$ is hence 
\begin{eqnarray}
c_{2+} & = & \sqrt{\frac{\pi}{2}}u^{-1}(x_{\theta})\tau\nu_{0}\left(U^{\prime}(x_{\theta})\,\left(\frac{G}{1+i\omega\tau}+x_{\theta}\,\frac{H}{2+i\omega\tau}\right)\right.\label{eq:c2_plus_threshold}\\
 &  & \left.+U(x_{\theta})\,\left(-n(\omega)+\left(\frac{1}{2}x_{\theta}\frac{G}{1+i\omega\tau}+(\frac{1}{2}x_{\theta}^{2}+i\omega\tau)\,\frac{H}{2+i\omega\tau}\right)\right)\right).\nonumber 
\end{eqnarray}
An analog expression holds for $c_{1+}$, which is, however, not needed
in the following, because we just need a condition for the solvability.
As the function $V$ is absent in the lower interval $x<x_{R}$, the
boundary condition at reset also determines $c_{2+}$, as seen in
the following. Expressing the solution in terms of $U$ and $V$
and subtracting the solutions above and below $x_{R}$, leads to the
linear system of equations

\[
\begin{array}{ccccccc}
U(x_{R}) & (c_{1+}-c_{1-}) & + & V(x_{R}) & c_{2+} & = & \left.q_{1}^{h}\right|_{x_{R}}\\
U^{\prime}(x_{R}) & (c_{1+}-c_{1-}) & + & V^{\prime}(x_{R}) & c_{2+} & = & \left.\partial_{x}q_{1}^{h}\right|_{x_{R}}.
\end{array}
\]
The coefficients $c_{1+}-c_{1-}$ and $c_{2+}$ are determined as
above as the solution of this system of linear equations
\begin{eqnarray}
\left(\begin{array}{c}
c_{1+}-c_{1-}\\
c_{2+}
\end{array}\right) & = & \frac{1}{W(x_{R})}\left(\begin{array}{cc}
V^{\prime} & -V\\
-U^{\prime} & U
\end{array}\right)\left(\begin{array}{c}
\left.q_{1}^{h}\right|_{x_{R}}\\
\partial_{x}\left.q_{1}^{h}\right|_{x_{R}}
\end{array}\right).\label{eq:Solution_UVsystem_reset}
\end{eqnarray}
Using the Wronskian $W(x_{R})=\sqrt{\frac{2}{\pi}}$ and the expressions
\prettyref{eq:threshold_cond} and \prettyref{eq:deriv_qh-1} for
the boundary values we obtain

\begin{eqnarray}
c_{2+} & = & \sqrt{\frac{\pi}{2}}u^{-1}(x_{R})\tau\nu_{0}\left(U^{\prime}(x_{R})\,\left(\frac{G}{1+i\omega\tau}+x_{R}\,\frac{H}{2+i\omega\tau}\right)\right.\label{eq:c2_plus_reset}\\
 &  & \left.+U(x_{R})\,\left(-n(\omega)+\left(\frac{1}{2}x_{R}\,\frac{G}{1+i\omega\tau}+(\frac{1}{2}x_{R}+i\omega\tau)\,\frac{H}{2+i\omega\tau}\right)\right)\right).\nonumber 
\end{eqnarray}
Equating \prettyref{eq:c2_plus_threshold} and \prettyref{eq:c2_plus_reset}
determines the transfer function

\begin{eqnarray}
n(\omega) & = & \frac{G}{1+i\omega\tau}\,\frac{\left.u^{-1}\,\left(U^{\prime}+\frac{x}{2}U\right)\right|_{x_{\theta}}^{x_{R}}}{\left.u^{-1}U\right|_{x_{\theta}}^{x_{R}}}\label{eq:iaf_transfer}\\
 & + & \frac{H}{2+i\omega\tau}\,\frac{\left.u^{-1}\,\left(x\,U^{\prime}+(\frac{x^{2}}{2}+i\omega\tau)\,U\right)\right|_{x_{\theta}}^{x_{R}}}{\left.u^{-1}U\right|_{x_{\theta}}^{x_{R}}}.\nonumber 
\end{eqnarray}
With the definition (using $m=i\omega\tau-\frac{1}{2}$ as defined
in \prettyref{eq:parabolic_cylinder_diffeq}) 
\begin{eqnarray}
\Phi_{\omega}(x) & \equiv\Phi(m,x)= & u^{-1}(x)\,U(m,x)\label{eq:Phi_notation}
\end{eqnarray}
follows 
\begin{eqnarray}
\Phi{}_{\omega}^{\prime}(x) & = & u^{-1}(x)(U^{\prime}(m,x)+\frac{x}{2}U(m,x))\label{eq:phi_primes}\\
\text{and}\nonumber \\
\Phi_{\omega}^{\prime\prime}(x) & = & \frac{x}{2}u^{-1}(U^{\prime}+\frac{x}{2}U)+u^{-1}(U^{\prime\prime}+\frac{1}{2}U+\frac{x}{2}U^{\prime})\nonumber \\
 & \stackrel{\eqref{eq:bl_solution}}{=} & \frac{x}{2}u^{-1}(U^{\prime}+\frac{x}{2}U)+u^{-1}((\frac{1}{4}x^{2}+i\omega\tau-\frac{1}{2})U+\frac{1}{2}U+\frac{x}{2}U^{\prime})\nonumber \\
 & = & u^{-1}\,(xU^{\prime}+(\frac{x^{2}}{2}+i\omega\tau)U).\nonumber 
\end{eqnarray}
 Inserting into \prettyref{eq:iaf_transfer} we obtain the final
result

\begin{eqnarray}
n(\omega) & = & \frac{G}{1+i\omega\tau}\,\frac{\left.\Phi_{\omega}^{\prime}(x)\right|_{x_{\theta}}^{x_{R}}}{\left.\Phi_{\omega}(x)\right|_{x_{\theta}}^{x_{R}}}\label{eq:transfer_final}\\
 & + & H\,\left(\frac{1}{2+i\omega\tau}\,\frac{\left.\Phi_{\omega}^{\prime\prime}(x)\right|_{x_{\theta}}^{x_{R}}}{\left.\Phi_{\omega}(x)\right|_{x_{\theta}}^{x_{R}}}\right).\nonumber 
\end{eqnarray}
This is the transfer function of the LIF model neuron as it was derived
in \citep{Brunel99,Lindner01_2934,Brunel01_2186}.

\subsection{Colored noise transfer function \label{sub:Colored-noise-transfer}}

We now consider the periodic modulation \prettyref{eq:mean_modulation}
of the mean input $\mu$ in the colored noise system \prettyref{eq:diff_LIF}.
Note that here we consider a modulation of $V$. If one is interested
in the linear response of the system with respect to a perturbation
of $I$, as it appears in the neural context due to synaptic input,
one needs to take into account the additional low pass filtering $\propto(1+i\omega\tau_{s})^{-1}$,
which is trivial. The modulated Fokker-Planck equation follows with
$f(y,s)=-y+\frac{\epsilon\mu}{\sigma}e^{i\omega\tau s}$ from \prettyref{eq:FP_2D},
and the effective system takes the form \prettyref{eq:perturbed_FP}
with $H=0$ (since we have no $\sigma-$modulation) and with specific
boundary conditions. We only consider a modulation of the mean $\mu$,
which dominates the response properties. The treatment of an additional
modulation of the variance $\sigma$ is shown in \citep{Schuecker14_arxiv_1411}
. The boundary conditions follow considering again the perturbation
ansatz for $\nu=\nu_{0}+k\nu_{1}+O(k^{2})$ which must hold for each
order of $k$ separately, i.e. 
\begin{eqnarray}
\nu(t) & = & \nu\,(1+n_{\mathrm{cn}}(\omega)\,e^{i\omega t}),\label{eq:ansatz_cn_tf}
\end{eqnarray}
so that to lowest order $k^{0}$ we have 
\begin{eqnarray}
\nu_{y}^{(0)}(\theta,s) & = & \nu_{y}^{(0)}\,(1+n_{\mathrm{cn}}(\omega)\,e^{i\omega\tau s}),\label{eq:nu_0_time_dep-1-1-1}
\end{eqnarray}
where $n_{\mathrm{cn}}$ is the colored noise transfer function to
be determined. Likewise to the case without synaptic filtering (\prettyref{sub:Perturbative-treatment-of})
we make a perturbative ansatz for the effective density 
\begin{eqnarray*}
\tilde{P}(y,s) & = & \tilde{P}(y)+e^{i\omega\tau s}\tilde{\hat{P}}(y).
\end{eqnarray*}
Therefore it follows from \prettyref{eq:Boundary_conditions_static}
\begin{eqnarray*}
\left.\tilde{P}(y,s)+e^{i\omega\tau s}\tilde{\hat{P}}(y)\right|_{\{R,\theta\}} & = & k\alpha\,\nu_{y}^{(0)}(1+n_{\mathrm{cn}}(\omega)\,e^{i\omega\tau s})
\end{eqnarray*}
and thus the boundary value for the time modulated part of the density
is 
\begin{eqnarray}
\left.\tilde{\hat{P}}(y)\right|_{\{R,\theta\}} & = & k\alpha\nu_{y}^{(0)}n_{\mathrm{cn}}(\omega).\label{eq:bc_mod_density}
\end{eqnarray}
In the white noise derivation (\prettyref{sub:White-noise-transfer-function})
we obtain the simultaneous boundary conditions for the homogeneous
part of the modulated density and its derivative. In the following
we adapt these conditions respecting the new boundary conditions of
the colored case \prettyref{eq:bc_mod_density} and perform the subsequent
steps of the derivation analogously to the white noise scenario. This
leads to an analytical expression for the transfer function valid
for synaptic filtering with small time constants $\tau_{s}$. Note
that with \prettyref{sub:Shifted-reset-and} we would directly obtain
an approximation for the colored noise transfer function $n_{\mathrm{cn}}$,
replacing $x_{\{R,\theta\}}\rightarrow x_{\{\tilde{R},\tilde{\theta}\}}$
in the white noise solution \prettyref{eq:transfer_final}, which
we denote by $\tilde{n}$. We will later show that the expression
obtained with the time-dependent modified boundary condition \prettyref{eq:bc_mod_density}
is to first order equivalent to$\tilde{n}$.

\subsubsection{Colored noise boundary condition for the modulated density}

The boundary condition for the function value of $q_{1}(x)=\frac{1}{\sqrt{2}}\tilde{\hat{P}}(x/\sqrt{2})u^{-1}(x)$
follows from \prettyref{eq:bc_mod_density} so

\begin{eqnarray}
q_{1}(x)\vert_{\{x_{R},x_{\theta}\}}=q_{1}^{h}+q_{1}^{p}\vert_{\{x_{R},x_{\theta}\}}= & Au^{-1}n_{\mathrm{cn}}(\omega).\label{eq: Boundary function value}
\end{eqnarray}
From here on we skip the dependence of $u$ on $\{x_{R},x_{\theta}\}$.
The contribution of the particular solution yields boundary conditions
for the homogeneous solution. With $H=0$ the particular solution
is \prettyref{eq:particular_solution}

\begin{eqnarray}
q_{1}^{p} & = & \frac{G}{1+i\omega\tau}a^{\dagger}q_{0}.\label{eq:particular}
\end{eqnarray}
The contribution of $a^{\dagger}q_{0}|_{\{x_{R},x_{\theta}\}}$ is

\begin{eqnarray}
a^{\dagger}q_{0}|_{\{x_{R},x_{\theta}\}} & = & (x-a)q_{0}|_{\{x_{R},x_{\theta}\}}\nonumber \\
 & = & \{x_{R},x_{\theta}\}Au^{-1}+\tau\nu\,u^{-1},\label{eq:a^+_q}
\end{eqnarray}
where we use \prettyref{eq:stationary_flux-1} and \prettyref{eq:Boundary_values_C.1-1-1}.
From \prettyref{eq: Boundary function value} we obtain the boundary
value of the homogeneous solution

\begin{eqnarray}
-q_{1}^{h}|_{\{x_{R},x_{\theta}\}} & = & q_{1}^{p}|_{\{x_{R},x_{\theta}\}}-Au^{-1}n_{\mathrm{cn}}(\omega)\nonumber \\
 & = & \frac{G}{1+i\omega\tau}(\{x_{R},x_{\theta}\}Au^{-1}+\tau\nu\,u^{-1})-Au^{-1}n_{\mathrm{cn}}(\omega).\label{eq:homogenous_boundary_value}
\end{eqnarray}

\subsubsection{Colored noise boundary condition for the derivative of the density}

From \prettyref{eq:flux_conditions} we have with $H=0$
\begin{eqnarray}
\tau\nu\,n_{\mathrm{cn}}(\omega) & = & S_{0}uq_{1}+S_{1}uq_{0}|_{\{x_{R},x_{\theta}\}}\nonumber \\
 & = & -uaq_{1}+uGq_{0}|_{\{x_{R},x_{\theta}\}}\label{eq: Boundary derivative}\\
 & = & -ua(q_{1}^{h}+q_{1}^{p})+uGq_{0}|_{\{x_{R},x_{\theta}\}}.\nonumber 
\end{eqnarray}
Here we again employ that $n_{\mathrm{cn}}(\omega)$ is simultaneously
valid for all orders of $k$. Therefore $\nu$, containing the first
order correction in $k$, appears on the left hand side. The contribution
of the particular solution is given by \prettyref{eq:flux_contribution_particular}
with $H=0$

\begin{eqnarray*}
-ua\,q_{1}^{p} & = & -u\left(\frac{G}{1+i\omega\tau}\right)q_{0}.
\end{eqnarray*}
Inserting in \prettyref{eq: Boundary derivative} and using \prettyref{eq:Boundary_values_C.1-1-1}
yields

\begin{eqnarray*}
\tau\nu\,n_{\mathrm{cn}}(\omega) & = & -\frac{GA}{1+i\omega\tau}+GA-uaq_{1}^{h}|_{\{x_{R},x_{\theta}\}}.
\end{eqnarray*}
Substituting $a=\frac{1}{2}x+\partial_{x}$ and the expression for
the function value \prettyref{eq:homogenous_boundary_value} this
expands to

\begin{eqnarray*}
\tau\nu\,n_{\mathrm{cn}}(\omega) & = & -\frac{GA}{1+i\omega\tau}+GA\\
 &  & +\frac{1}{2}\{x_{R},x_{\theta}\}\left(\frac{G}{1+i\omega\tau}(\{x_{R},x_{\theta}\}A+\tau\nu)\right)\\
 &  & -\frac{1}{2}\{x_{R},x_{\theta}\}A\,n_{\mathrm{cn}}(\omega)\\
 &  & -u\partial_{x}q_{1}^{h}\vert_{\{x_{R},x_{\theta}\}}.
\end{eqnarray*}
The terms in the first line of the right hand side can be simplified
by elementary algebraic manipulations so that rearranging for the
derivative yields

\begin{eqnarray}
u\partial_{x}q_{1}^{h}\vert_{\{x_{R},x_{\theta}\}} & = & -n_{\mathrm{cn}}(\omega)(\tau\nu+\frac{1}{2}\{x_{R},x_{\theta}\}A)\nonumber \\
 &  & +\frac{G}{1+i\omega\tau}\left(i\omega\tau A+\frac{1}{2}\{x_{R},x_{\theta}\}(\{x_{R},x_{\theta}\}A+\tau\nu)\right).\label{eq:homogeneous boundary derivative}
\end{eqnarray}

\subsubsection{Solvability condition}

Above we have determined the boundary conditions for the function
value \prettyref{eq:homogenous_boundary_value} as well as the derivative
\prettyref{eq:homogeneous boundary derivative}. Now we consider the
solvability condition to determine the transfer function as in \prettyref{sub:Solvability_delta}.
According to \prettyref{eq:Solution_UVsystem} and \prettyref{eq:Solution_UVsystem_reset}
we obtain two equations determining the coefficient $c_{2+}$ of the
homogeneous solution in \prettyref{eq:general_solution_q1} for the
conditions at $x_{\theta}$ and $x_{R}$ respectively 
\begin{eqnarray*}
c_{2+}^{\theta} & = & \sqrt{\frac{\pi}{2}}[-U^{\prime}q_{1}^{h}(x_{\theta})+U\partial_{x}q_{1}^{h}(x_{\theta})]\\
c_{2+}^{R} & = & \sqrt{\frac{\pi}{2}}[-U^{\prime}\left.q_{1}^{h}\right|_{x_{R}}+U\left.\partial_{x}q_{1}^{h}\right|_{x_{R}}].
\end{eqnarray*}
 Since the two coefficients $c_{2+}^{\theta}$ and $c_{2+}^{R}$ must
be equal, the transfer function $n_{\mathrm{cn}}(\omega)$ is determined
by $c_{2+}^{\theta}=c_{2+}^{R}$. Inserting \prettyref{eq:homogenous_boundary_value}
and \prettyref{eq:homogeneous boundary derivative} we sort for terms
proportional to $n_{\mathrm{cn}}(\omega)$ and obtain 
\begin{eqnarray}
 &  & \left.A\left(U^{\prime}u^{-1}+Uu^{-1}\frac{1}{2}x\right)+Uu^{-1}\tau\nu\right|_{x_{R}}^{x_{\theta}}n_{\mathrm{cn}}(\omega)\nonumber \\
 & = & \frac{G}{1+i\omega\tau}\left.\tau\nu u^{-1}\left(U^{\prime}+\frac{1}{2}xU\right)\right|_{x_{R}}^{x_{\theta}}\label{eq:transfer_function_complete}\\
 & + & \frac{AG}{1+i\omega\tau}\left.u^{-1}\left(U^{\prime}x+U(i\omega\tau+\frac{1}{2}x^{2})\right)\right|_{x_{R}}^{x_{\theta}}.\nonumber 
\end{eqnarray}
Using \prettyref{eq:Phi_notation} and \prettyref{eq:phi_primes}
we can write the transfer function as
\begin{eqnarray}
n_{\mathrm{cn}}(\omega) & = & \tau\nu\,\frac{G}{1+i\omega\tau}\,\frac{\Phi_{\omega}^{\prime}\vert_{x_{\theta}}^{x_{R}}}{A\Phi_{\omega}^{\prime}+\Phi_{\omega}\tau\nu\vert_{x_{\theta}}^{x_{R}}}\nonumber \\
 &  & +\frac{AG}{1+i\omega\tau}\,\frac{\Phi_{\omega}^{\prime\prime}\vert_{x_{\theta}}^{x_{R}}}{A\Phi_{\omega}^{\prime}+\Phi_{\omega}\tau\nu\vert_{x_{\theta}}^{x_{R}}}.\label{eq:transfer_function_nonlin}
\end{eqnarray}

\subsubsection{Linearization in $k$}

Since we neglected all terms of second order in $k$ we state our
final result linearly in $k$ and neglect higher orders by performing
an expansion into a geometric series. From \prettyref{eq:transfer_function_nonlin}
we obtain 
\begin{eqnarray}
n_{\mathrm{cn}}(\omega) & = & \frac{G}{1+i\omega\tau}\left[\frac{\Phi_{\omega}^{\prime}\vert_{x_{\theta}}^{x_{R}}}{\Phi_{\omega}\vert_{x_{\theta}}^{x_{R}}}+\frac{k\alpha\,\nu_{0}}{\sqrt{2}\nu}\left(\frac{\Phi_{\omega}^{\prime\prime}\vert_{x_{\theta}}^{x_{R}}}{\Phi_{\omega}\vert_{x_{\theta}}^{x_{R}}}-\left(\frac{\Phi_{\omega}^{\prime}\vert_{x_{\theta}}^{x_{R}}}{\Phi_{\omega}\vert_{x_{\theta}}^{x_{R}}}\right)^{2}\right)\right].\label{eq:Transfer_colored_lin}
\end{eqnarray}
The first term is denoted by $n_{\mathrm{cn}}^{\mathrm{wn}}$ since
it is equivalent to the white noise solution \prettyref{eq:transfer_final}.
As mentioned earlier the correction terms in $k$ could be obtained
by a shift in the boundaries in $n_{\mathrm{cn}}^{\mathrm{wn}}$
\begin{eqnarray*}
\tilde{x}_{R} & = & x_{R}+\frac{\alpha}{\sqrt{2}}k\\
\tilde{x}_{\theta} & = & x_{\theta}+\frac{\alpha}{\sqrt{2}}k
\end{eqnarray*}
and a Taylor expansion in $k$

\begin{eqnarray*}
n_{\mathrm{cn}}(\omega) & = & \frac{G}{1+i\omega\tau}\frac{\Phi_{\omega}^{\prime}\vert_{\tilde{x}_{\theta}}^{\tilde{x}_{R}}}{\Phi_{\omega}\vert_{\tilde{x}_{\theta}}^{\tilde{x}_{R}}}\\
 & \stackrel{\textmd{quotient rule}}{=} & \frac{G}{1+i\omega\tau}\left[\frac{\Phi_{\omega}^{\prime}\vert_{x_{\theta}}^{x_{R}}}{\Phi_{\omega}\vert_{x_{\theta}}^{x_{R}}}+\frac{k\alpha}{\sqrt{2}}\left(\frac{\Phi_{\omega}^{\prime\prime}\vert_{x_{\theta}}^{x_{R}}}{\Phi_{\omega}\vert_{x_{\theta}}^{x_{R}}}-\left(\frac{\Phi_{\omega}^{\prime}\vert_{x_{\theta}}^{x_{R}}}{\Phi_{\omega}\vert_{x_{\theta}}^{x_{R}}}\right)^{2}\right)\right],
\end{eqnarray*}
which is to first order equivalent to \prettyref{eq:Transfer_colored_lin}
since $\frac{\nu_{0}}{\nu}=\frac{\nu_{0}}{\nu_{0}+k\nu_{1}}=1-k\frac{\nu_{1}}{\nu_{0}}+O(k^{2})$.

\section{Numerical Simulation}

We perform direct simulations of \prettyref{eq:diff_LIF} with periodic
modulation of the mean \prettyref{eq:mean_modulation}. Simulations
were done in NEST \citep{Gewaltig_07_11204}. For each data point
in \prettyref{fig: contributions} we simulate $10000\ms$, whereby
we allow for a warm-up time of $100\ms$, and average over $100000$
neurons (model: ``iaf\_psc\_exp''). We use a time resolution of
$dt=0.001\ms$ to have a good agreement to the analytical limit $dt\rightarrow0$,
which is important regarding the implementation of the white noise
$\xi$ as a step-wise constant current with stepsize $dt$. We perform
a fast Fourier transform on the summed spike trains to obtain the
amplitude and phase of the transfer function.

\section{Discussion}

\begin{figure}
\centering{}\includegraphics{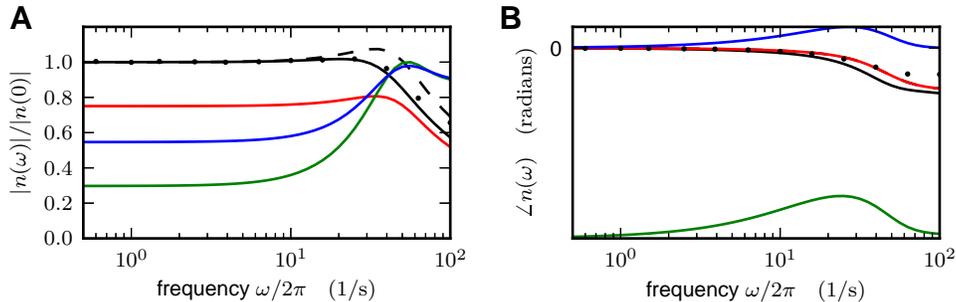}\caption{Absolute value (\textbf{A}) and phase (\textbf{B}) of the transfer
function for $\mu=18.94\protect\mV$, $\sigma=1.5\protect\mV$, $\theta=19.5\protect\mV$,
$V_{r}=14.5\protect\mV$, $\tau_{m}=10\protect\ms$, $\tau_{s}=1.0\protect\ms$.
Different contributions $n_{\mathrm{cn}}^{\mathrm{wn}}$ (red), $n_{\mathrm{cn}}^{\mathrm{H}}$
(blue) and $n_{\mathrm{cn}}^{\mathrm{square}}$ (green) and complete
solution $\tilde{n}_{\mathrm{cn}}$ (black) compared to white noise
case (dashed). Complete transfer functions for white noise and colored
noise separately normalized to zero frequency limit $n(0)=\frac{d\nu_{0}}{d\mu}$
and $n_{cn}(0)=\frac{d\nu}{d\mu}$ (cf. \prettyref{app:Zero-frequency-limit}).\label{fig: contributions}}
\end{figure}

Finally we are in the position to compare the colored noise transfer
function \prettyref{eq:Transfer_colored_lin} to the white noise case
\prettyref{eq:transfer_final}. To this end we examine the contributions
of the different terms in \prettyref{eq:Transfer_colored_lin}
\begin{eqnarray*}
n_{\mathrm{cn}}(\omega) & = & \underbrace{\frac{G}{1+i\omega\tau}\frac{\Phi_{\omega}^{\prime}\vert_{x_{\theta}}^{x_{R}}}{\Phi_{\omega}\vert_{x_{\theta}}^{x_{R}}}}_{n_{\mathrm{cn}}^{\mathrm{wn}}}+\underbrace{\frac{G}{1+i\omega\tau}\frac{k\alpha\,\nu_{0}}{\sqrt{2}\nu}\frac{\Phi_{\omega}^{\prime\prime}\vert_{x_{\theta}}^{x_{R}}}{\Phi_{\omega}\vert_{x_{\theta}}^{x_{R}}}}_{n_{\mathrm{cn}}^{H}}-\underbrace{\frac{G}{1+i\omega\tau}\frac{k\alpha\,\nu_{0}}{\sqrt{2}\nu}\left(\frac{\Phi_{\omega}^{\prime}\vert_{x_{\theta}}^{x_{R}}}{\Phi_{\omega}\vert_{x_{\theta}}^{x_{R}}}\right)^{2}}_{n_{\mathrm{cn}}^{\mathrm{square}}},
\end{eqnarray*}
shown in \prettyref{fig: contributions}. We notice that the correction
term $n_{\mathrm{cn}}^{H}$ is similar to the $H-$term in \eqref{eq:transfer_final},
meaning that colored noise has a similar effect on the transfer function
as a modulation of the variance in the white noise case. For infinite
frequencies this similarity was already found: modulation of the variance
leads to finite transmission at infinite frequencies in the white
noise system \citep{Lindner01_2934} and the same is true for modulation
of the mean in the presence of filtered noise \citep{Brunel01_2186}.
The latter can be calculated in the two-dimensional Fokker-Planck
problem and the result is (cf. \prettyref{app:HF_limit}) 
\begin{eqnarray*}
n_{\mathrm{cn}}^{\mathrm{lim,2D}}(\omega) & = & 1.3238\,\frac{\epsilon k\mu}{\sigma}.
\end{eqnarray*}
However, our analytical expression behaves differently and does not
provide an accurate limit. The two correction terms $n_{\mathrm{cn}}^{\mathrm{square}}$
and $n_{\mathrm{cn}}^{H}$ cancel each other (see \prettyref{fig: contributions}),
since \citep[12.8]{DLMF12}
\begin{eqnarray*}
\Phi^{\prime}(i\omega\tau-\frac{1}{2},x) & = & -i\omega\tau\,\Phi(i\omega\tau+\frac{1}{2},x)\\
\Phi^{\prime\prime}(i\omega\tau-\frac{1}{2},x) & = & i\omega\tau(i\omega\tau+1)\,\Phi(i\omega\tau+\frac{3}{2},x),
\end{eqnarray*}
so $\Phi^{\prime\prime}\rightarrow(i\omega\tau)^{2}\,\Phi$ and $\Phi^{\prime^{2}}\rightarrow(i\omega\tau)^{2}\,\Phi^{2}$.
Thus $n_{\mathrm{cn}}^{\mathrm{wn}}$ is the only term left, meaning
that the transfer function decays to zero as in the white noise case.
This discrepancy originates from our derivation of the boundary value
\prettyref{eq: Boundary function value}: We neglect all terms with
time derivatives in \prettyref{eq:FP_time_depending_P}, since they
are of second and third order in $k$ and we assume $\omega\tau k\ll1$,
although this holds only true for moderate frequencies. Note that
in fact only the terms including time derivatives in \prettyref{eq:FP_time_depending_P}
play a role in the limit $\omega\rightarrow\infty$ as seen in Eq.
\prettyref{eq:Phat_dzP}, leading to the correct limit in the two-dimensional
system. We also expect that the deviations at high frequencies increase
with the synaptic time constant, since the neglected terms are $\propto\omega k\propto\omega\sqrt{\tau_{s}}$.

Nevertheless up to moderate frequencies the analytical expression
for the transfer function found in the present work \prettyref{eq:Transfer_colored_lin}
is in agreement to direct simulations as shown in \prettyref{fig: contributions}.
In this regime the color of the noise suppresses the resonant peak
and reduces the cutoff frequency compared to the white noise case.
The effect of the noise at intermediate frequencies is hence opposite
to its effect in the high frequency limit \citep{Brunel01_2186}.
This constitutes a novel insight into the dependence of the transfer
properties of LIF neurons on the details of synaptic dynamics.

In summary, in these pages we supply intermediate steps of the derivation
of the theory presented in \citep{Schuecker14_arxiv_1411} that may
prove useful for the reader interested in further developments of
the approach. Showing the applicability of the alternative method
of reduction of a colored-noise to a white noise system, namely explicitly
taking into account the time-dependent boundary condition, may facilitate
future extensions of the theory that go beyond the first order in
the perturbation parameter $k$. We hope that the pedagogical presentation
chosen here, exposing the analogy of the LIF model to the quantum
harmonic oscillator and including all intermediate steps of the calculations,
will be of use especially for students entering the field with a general
background in physics, math or equivalent, but unfamiliar with Fokker-Planck
theory. 

\pagebreak{}

\appendix

\section{Autocorrelation of $z$\label{app:auto1d}}

Fourier transformation of the second equation in \eqref{eq:diffeq_general}
yields 

\begin{eqnarray*}
\hat{z} & = & \frac{k}{1+i\omega k^{2}}\xi(\omega),
\end{eqnarray*}
where $\hat{z}$ denotes the Fourier transform of $z$. The power-spectrum
then follows as
\begin{eqnarray*}
\langle\hat{z}(\omega)\hat{z}(-\omega)\rangle & = & k^{2}\frac{1}{(1+i\omega k^{2})(1-i\omega k^{2})},
\end{eqnarray*}
where we used $\langle\xi(\omega)\xi(-\omega)\rangle=1$. We perform
the back transform using the residue theorem and assume $s^{\prime}>0$
(which allows us closing the contour in the upper complex half plane
due to the term $e^{is^{\prime}\omega}$), thus

\begin{eqnarray*}
\langle z(s)z(s+s^{\prime})\rangle & = & \frac{1}{2\pi}\int k^{2}\frac{e^{i\omega s^{\prime}}}{(1+i\omega k^{2})(1-i\omega k^{2})}\,d\omega\\
 & \stackrel{\mathbf{\mathrm{\mathrm{\text{\ensuremath{\mathrm{\text{pole }\omega=\frac{i}{k^{2}}}}}}}}}{=} & k^{2}\frac{1}{2\pi}2\pi i\frac{(\omega-\frac{i}{k^{2}})e^{-s^{\prime}/k^{2}}}{ik^{2}(\omega-\frac{i}{k^{2}})(1+1)}\\
 & = & \frac{e^{-s^{\prime}/k^{2}}}{2}.
\end{eqnarray*}
If $s^{\prime}<0$ we have to close the contour in the lower half
plane. Together we get
\begin{eqnarray*}
\langle z(s)z(s+s^{\prime})\rangle & = & \frac{e^{-|s^{\prime}|/k^{2}}}{2}.
\end{eqnarray*}

\section{Zero frequency limit\label{app:Zero-frequency-limit}}

The zero frequency limit of the colored noise transfer function $n_{\mathrm{cn}}(\omega)$
is given by the derivative of the firing rate \eqref{eq:firing_rate_final}
with respect to $\mu$. For brevity we introduce $\Psi(x)=u^{-2}F=e^{x^{2}/2}\left(\sqrt{\frac{\pi}{2}}(1+\erf(\frac{x}{\sqrt{2}})\right)$
and $S=\left(\int_{x_{R}}^{x_{\theta}}\Psi\,dx\right)$ so that 

\begin{eqnarray*}
\tau\nu & = & \tau\nu_{0}-\frac{\alpha k}{\sqrt{2}}\frac{\left.\Psi\right|_{x_{R}}^{x_{\theta}}}{S^{2}}.
\end{eqnarray*}
With $x=\sqrt{2}\frac{V-\mu}{\sigma}$ we have

\begin{eqnarray*}
\frac{d\Psi(x)}{d\mu} & = & -\frac{\sqrt{2}}{\sigma}\sqrt{\frac{\pi}{2}}\left(xe^{x^{2}/2}(1+\erf(\frac{x}{\sqrt{2}}))+\sqrt{\frac{2}{\pi}}\right)\\
\text{\text{and}}\\
\frac{dS}{d\mu} & = & (-\frac{\sqrt{2}}{\sigma})(\Psi(x_{\theta})-\Psi(x_{R})),
\end{eqnarray*}
yielding 

\begin{eqnarray*}
\frac{d\nu}{d\mu} & = & \frac{d\nu_{0}}{d\mu}-\frac{\alpha k}{\sqrt{2}\tau}\frac{(\Psi^{\prime}(x_{\theta})-\Psi^{\prime}(x_{R}))S^{2}-(-\frac{\sqrt{2}}{\sigma})(\Psi(x_{\theta})-\Psi(x_{R}))2S((\Psi(x_{\theta})-\Psi(x_{R}))}{S^{4}}\\
 & = & \frac{d\nu_{0}}{d\mu}-\frac{\alpha k}{\sqrt{2}\tau}\frac{(\Psi^{\prime}(x_{\theta})-\Psi^{\prime}(x_{R}))S+\frac{2\sqrt{2}}{\sigma}(\Psi(x_{\theta})-\Psi(x_{R}))^{2}}{S^{3}}.
\end{eqnarray*}

\section{High frequency limit\label{app:HF_limit}}

For completeness we rederive the high frequency limit of the transfer
function in the two-dimensional Fokker-Planck problem, closely following
\citet{Brunel01_2186}. The firing rate is given by the probability
flux in $y$-direction at threshold, marginalized over $z$. With
the ansatz of a sinusoidal modulation of the density we therefore
have
\begin{eqnarray*}
\nu_{y}(\theta,s) & = & \int_{-\infty}^{\infty}\frac{1}{k}zP(y_{\theta},z,s)\,dz,\\
 & = & \int_{-\infty}^{\infty}\frac{1}{k}z(P(y_{\theta},z)+\hat{P}(y_{\theta},z)e^{i\omega\tau s})\,dz,
\end{eqnarray*}
with \prettyref{eq:ansatz_cn_tf} resulting in 
\begin{eqnarray}
\nu\tau\,n_{\mathrm{cn}}(\omega) & = & \int_{-\infty}^{\infty}\frac{1}{k}z\hat{P}(y_{\theta},z)\,dz.\label{eq:n_omega_limit}
\end{eqnarray}
Inserting $f(y,s)=-y+\frac{\epsilon\mu}{\sigma}e^{i\omega\tau s}$
in \eqref{eq:FP_time_depending_P} and using the perturbation ansatz
$P(y_{\theta},z,s)=P(y_{\theta},z)+\hat{P}(y_{\theta},z)e^{i\omega\tau s}$
we get for $\omega\rightarrow\infty$
\begin{eqnarray}
\hat{P}(y_{\theta},z) & = & -\epsilon k\frac{\mu}{\sigma}\partial_{z}P(y_{\theta},z),\label{eq:Phat_dzP}
\end{eqnarray}
which shows that the density is necessarily time-modulated up to arbitrary
high frequencies. Together with \eqref{eq:n_omega_limit} we have

\begin{eqnarray}
\tau\nu\,n_{\mathrm{cn}}(\omega) & = & -\int_{-\infty}^{\infty}\epsilon\frac{1}{k}zk\frac{\mu}{\sigma}\partial_{z}P(y_{\theta},z)\,dz\nonumber \\
 & \stackrel{\text{\text{perturbation series in k}}}{=} & -\int_{-\infty}^{\infty}\epsilon z\frac{\mu}{\sigma}\partial_{z}\left(Q^{(0)}(y_{\theta})+kQ^{(1)}(y_{\theta},z)\right)\frac{e^{-z^{2}}}{\sqrt{\pi}}\,dz\nonumber \\
 & \stackrel{Q^{(0)}(y_{\theta})=0}{=} & -\int_{-\infty}^{\infty}\epsilon z\frac{\mu}{\sigma}k\partial_{z}\left(Q^{(1)}(y_{\theta},z)\frac{e^{-z^{2}}}{\sqrt{\pi}}\right)\,dz\nonumber \\
 & \stackrel{\text{\text{integration by parts}}}{=} & \int_{-\infty}^{\infty}\epsilon\frac{\mu}{\sigma}kQ^{(1)}(y_{\theta},z)\frac{e^{-z^{2}}}{\sqrt{\pi}}\,dz\nonumber \\
 & \stackrel{\text{ \eqref{eq:bl_solution} and \ensuremath{r=0}}}{=} & \int_{-\infty}^{\infty}\epsilon\frac{\mu}{\sigma}k2\tau\nu_{0}\left(\frac{\alpha}{2}+z+\sum_{n=1}^{\infty}b_{n}(z)\right)\frac{e^{-z^{2}}}{\sqrt{\pi}}\,dz\mbox{}\nonumber \\
 & \stackrel{\text{symmetry and }\citeklosek}{=} & \epsilon k\frac{\mu}{\sigma}\tau\nu_{0}(\alpha-\frac{1}{\sqrt{2}}\sum_{n=1}^{\infty}\frac{N(\sqrt{n})}{n!\sqrt{n}}e^{-\frac{n}{2}}n^{\frac{n}{2}})\nonumber \\
 & \stackrel{\text{\citebrunel}}{\simeq} & 1.3238\,\epsilon k\frac{\mu}{\sigma}\tau\nu_{0}.\label{eq:HF_limit_1-1}
\end{eqnarray}
The numerical value in the last expression is taken from \citep{Fourcaud02}.
We divide by the firing rate in the colored-noise case \eqref{eq:firing_rate_final}
$\nu=\nu_{0}+k\nu_{1}$ and linearize the right hand side in $k$
which gives 

\begin{eqnarray}
n_{\mathrm{cn}}(\omega) & \simeq & \frac{1.3238\,\epsilon k\mu}{\sigma}\,\frac{\nu_{0}}{\nu_{0}+k\nu_{1}}\nonumber \\
 & = & \frac{1.3238\,\epsilon k\mu}{\sigma}(1-k\frac{\nu_{1}}{\nu_{0}})+O(k^{2})\nonumber \\
 & = & 1.3238\,\frac{\epsilon k\mu}{\sigma}+O(k^{2}).\label{eq:limit_FB}
\end{eqnarray}

\end{document}